\documentclass[letterpaper]{JHEP3}

\JHEPspecialurl{http://jhep.sissa.it/JOURNAL/JHEP3.tar.gz}

\usepackage{epsfig,multicol,bbm}
\usepackage{amsmath,amsfonts,amssymb,theorem,mathrsfs,times}
\usepackage{graphicx}
\usepackage{subfigure}

\newcommand\fverb{\setbox\fverbbox=\hbox\bgroup\verb}
\newcommand\fverbdo{\egroup\medskip\noindent%
			\fbox{\unhbox\fverbbox}\ }
\newcommand\fverbit{\egroup\item[\fbox{\unhbox\fverbbox}]}
\newbox\fverbbox

\title{Magnetic black holes with higher-order curvature and gauge corrections in even dimensions }

\author{Hideki Maeda\\
Centro de Estudios Cient\'{\i}ficos (CECS), Casilla 1469, Valdivia, Chile\\
E-mail: \email{hideki@cecs.cl}}
\author{Mokhtar Hassa\"{\i}ne\\
Instituto de Matem\'atica y F\'{\i}sica, Universidad de Talca, Casilla 747, Talca, Chile\\
E-mail: \email{hassaine@inst-mat.utalca.cl}}
\author{Cristi{\'a}n Mart\'{\i}nez\\
Centro de Estudios Cient\'{\i}ficos (CECS), Casilla 1469, Valdivia, Chile\\
Centro de Ingenier\'{\i}a de la Innovaci\'on del CECS (CIN),
Valdivia, Chile\\
E-mail: \email{martinez@cecs.cl}}

\received{\today} 		%%
\accepted{\today}		%% These are for published papers.

\newcommand{\ma}[1]{\mbox{$\mathcal{#1}$}}

\abstract{We obtain magnetic black-hole solutions in arbitrary
$n(\ge 4)$ even dimensions for an action given by the
Einstein-Gauss-Bonnet-Maxwell-$\Lambda$ pieces with the $F^4$
gauge-correction terms. This action arises in the low energy limit
of heterotic string theory with constant dilaton and vanishing higher form
fields. The spacetime is assumed to be a warped
product ${\ma M}^{2} \times {\ma K}^{n-2}$, where ${\ma K}^{n-2}$ is
a $(n-2)$-dimensional Einstein space satisfying a condition on its
Weyl tensor, originally considered by Dotti and Gleiser. Under
a few reasonable assumptions, we establish the generalized
Jebsen-Birkhoff theorem for the magnetic solution in the
case where the orbit of the warp factor on ${\ma K}^{n-2}$ is
non-null. We prove that such magnetic solutions do not exist in odd
dimensions. In contrast, in even dimensions, we obtain an explicit
solution in the case where ${\ma K}^{n-2}$ is a product manifold of
$(n-2)/2$ two-dimensional maximally symmetric spaces
with the same constant warp factors. In this latter case, we show
that the global structure of the spacetime sharply depends on the existence of the
gauge-correction terms as well as the number of spacetime
dimensions.}

\begin{document} 

%\maketitle  IS IGNORED %%%%%%%%%%%

\newpage
\section{Introduction}
What is the description of a black hole in the quantum theory of
gravity? The answer to this question is one of the ultimate goals of
modern physics. String theory which is consistently formulated in
higher dimensions is a promising candidate of a unified theory. In
string theory, the extra dimensions are usually considered to be
compactified, and as a result, the effect of string theory is
negligible for large astrophysical black holes. However, in order to
discuss the formation of tiny black holes in the upcoming Large
Hadron Collider (LHC) or the final fate of an evaporating black hole
by the Hawking radiation, the effect of string theory cannot be
neglected. If the horizon size becomes comparable to the curvature
radius of the compactified extra dimensions, the black hole
effectively becomes higher-dimensional. Although the
non-perturbative aspects of string theory have been intensively
investigated in recent years focusing on the conjecture of
anti-de~Sitter(AdS)/conformal field theory(CFT)
correspondence~\cite{ads/cft}, the full description of a black hole
in string theory is still far from complete.

Another possible approach to study the string effect is to study
black holes in the low-energy classical theory. Higher-dimensional
general relativity is realized as the lowest order in the Regge
slope expansion of strings. Then, it is known that the
higher-curvature terms appear as the next stringy compensation.
Among five types of string theories, there appears the so-called
Gauss-Bonnet term in the heterotic string case~\cite{Gross,gs1987},
which is a ghost-free and renormalizable combination of the
quadratic curvature terms~\cite{Zwiebach:1985uq}. The active study
of the Gauss-Bonnet black holes has its roots in the discovery of
the spherically symmetric vacuum Boulware-Deser-Wheeler
solution~\cite{bdw}, which is the counterpart of the Tangherlini
solution in general relativity~\cite{tangherlini1963}. However, in
spite of the considerable progress in recent years on this subject,
even the stationary axisymmetric rotating vacuum black-hole
solution, namely the counterpart of the Myers-Perry solution in
general relativity~\cite{mp1986}, has not been obtained yet.
(See~\cite{GB-review} for recent reviews.)

Since gauge fields are fundamental in the standard model, black
holes with gauge fields are also important from the viewpoint of
string theory. The Gauss-Bonnet black-hole solution with Maxwell electric
charge was obtained by Wiltshire~\cite{wiltshire1986} and has been
generalized to the topological case with a cosmological
constant~\cite{lorenz-petzold1988,EGBBH,cai2002}. Indeed, in the
low-energy limit of heterotic string theory, the higher-order
correction terms appear also for the Maxwell gauge
field~\cite{gs1987}. Therefore, in order to study the semi-classical
aspects of black holes, it is fair to consider not only the
correction terms arising from the gravity side but also those
related to the gauge field. This is one of the motivations of the
present paper. Here we will be concerned with the Einstein
equations supplemented by the Gauss-Bonnet term with a source
provided by the Maxwell field with the $F^4$ gauge-correction terms.
The motivation of considering such Lagrangian are multiple. Firstly,
the electrically charged Gauss-Bonnet black holes with the higher
$F$-terms is a current well-studied topic~\cite{kmp2007,ap2009,cns2008}. 
In Ref.~\cite{ap2009}, the effects
of the $F^4$ gauge-correction terms on the thermodynamical aspects
of black holes have been fully investigated. In addition, the
Lagrangian adopted in~\cite{ap2009} is considered as an interesting
model in the low-energy limit of heterotic string theory since
it contains the Lagrangian in the low-energy limit of the
ten-dimensional $E_8 \times E_8$ or $SO(32)$ heterotic string
theory. To be more complete, the action considered in \cite{ap2009}
also arises in four dimensions from the corrections to the
magnetically charged string black holes~\cite{natsuume1994} by
setting the dilaton to be constant.

In this paper, we are interested in magnetic
black holes in arbitrary $n (\ge 4)$ dimensions. As is well-known,
the Maxwell electromagnetic field is a fundamental gauge field in
physics which presents some attractive features in four dimensions.
Among other things, the electro-magnetic 2-form duality as well as
the conformal invariance of the Maxwell action are effective only in four dimensions. 
In higher dimensions, maybe because of the lack of these
properties, the Maxwell field is not well tamed. As an appealing
example to illustrate this fact, the higher-dimensional version of
the Kerr-Newman solution has not been obtained in general relativity so far. (See~\cite{iu2003}
for discussions.) Another example is provided with the study of
magnetic black holes. Indeed, even if some results are
known~\cite{magnetic}, the problem of finding such solutions even in
the case of spacetimes locally ${\ma M}^{2} \times {\ma K}^{n-2}$ is
still an open problem.
(Here ${\ma K}^{n-2}$ is a $(n-2)$-dimensional
Einstein space.) In this case, the difficulties may arise because
the number of the magnetic components of the Faraday tensor (in
contrast with the standard electric solution) grows with the spacetime
dimensions.

In general relativity, it is well-known that replacing the
$(n-2)$-dimensional space of positive constant curvature in the
Schwarzschild-Tangherlini spacetime by {\it any} $(n-2)$-dimensional
Einstein space with positive curvature will still provide a solution
of the vacuum Einstein equations. However, this is not the case in
the presence of the Gauss-Bonnet term. The reason is that, unlike
the Einstein tensor, the quadratic Gauss-Bonnet tensor contains the
Riemann tensors explicitly, and hence it gives a more severe
constraint on the $(n-2)$-dimensional Einstein space. In the
Boulware-Deser-Wheeler vacuum spacetime, ${\ma K}^{n-2}$ is
maximally symmetric, namely a $(n-2)$-dimensional space of positive constant
curvature. Considering a $(n-2)$-dimensional Einstein space for
${\ma K}^{n-2}$ in Einstein-Gauss-Bonnet gravity, Dotti and Gleiser
derived a consistency condition on the Weyl tensor on ${\ma
K}^{n-2}$ with the field equations and obtained an exact vacuum black-hole
solution~\cite{dg2005}. The effect of the Weyl tensor appears in the
metric function and makes the spacetime geometry
quite non-trivial. 
In this paper, we will consider the Dotti-Gleiser
condition as an assumption in order to obtain magnetic
black-hole solutions in the case where ${\ma K}^{n-2}$ is an Einstein
space.
(Both static and dynamical aspects of Gauss-Bonnet black holes with this class of non-constant curvature horizons have been recently studied in~\cite{maeda2010}.)

The plan of the paper is organized as follows. In the next section,
we present the model and clearly state our assumptions. In section
III, we obtain the unique possible form of the metric compatible
with the magnetic field and we will prove the non-existence of the
magnetic solutions in odd dimensions. We also derive explicit
solutions in even dimensions. In the section IV, we discuss the
properties of the solution and show that the black-hole
configurations arise for a particular range of the parameters. In
the section V, we summarize our results.
Our basic notations follow \cite{wald}. The
conventions of curvature tensors are $[\nabla _\rho
,\nabla_\sigma]V^\mu ={R^\mu }_{\nu\rho\sigma}V^\nu$ and $R_{\mu \nu
}={R^\rho }_{\mu \rho \nu }$, where $\nabla_\nu$ is the covariant
derivative. The Minkowski metric is taken to be the mostly plus
sign, and Roman indices run over all spacetime indices. We adopt the
units in which only the $n$-dimensional gravitational constant $G_n$
is retained.

%======================================%
%<<<<<<<<<<<< SECTION I  >>>>>>>>>>>>>>%
%======================================%

%%%%%%%%%%%%%%%%%%%%%%%%%%%%%%%%%%%%%%%%%%%%%%%%%%%%%%%%%%%%%%%%%%%%%%%%%%%
\section{Einstein-Gauss-Bonnet-Maxwell-$\Lambda$ system with gauge-correction terms}
%%%%%%%%%%%%%%%%%%%%%%%%%%%%%%%%%%%%%%%%%%%%%%%%%%%%%%%%%%%%%%%%%%%%%%%%%%%
In this section we consider the Einstein action supplemented by the
cosmological constant and the Gauss-Bonnet term in arbitrary
dimensions. The matter source is provided by the Maxwell action and the $F^4$ gauge-corrections terms built up
with the Faraday tensor. After deriving the field equations and
explaining the origin of such an action, we will assume that the
spacetime geometry is given by a warped product ${\ma M}^{2} \times
{\ma K}^{n-2}$, where ${\ma K}^{n-2}$ is a $(n-2)$-dimensional
Einstein space satisfying a certain condition presented below.

\subsection{Preliminaries}
In arbitrary dimensions $n(\ge 4)$, we consider the following action
%\begin{widetext}
\begin{eqnarray}
\label{action}
S&=&S_{\rm gravity}+S_{\rm matter},\\
S_{\rm gravity}&=&\int d^nx\sqrt{-\det(g_{\mu\nu})}\biggl[\frac{1}{2\kappa_n^2}(R-2\Lambda+\alpha{L}_{GB}) \biggr],\\
S_{\rm matter}&=&-\frac{1}{4g^2}\int d^nx\sqrt{-\det(g_{\mu\nu})}F_{\mu\nu}F^{\mu\nu} \nonumber \\
&+&\int
d^nx\sqrt{-\det(g_{\mu\nu})}\biggl[c_1(F_{\mu\nu}F^{\mu\nu})^2+c_2F_{\mu\nu}F^{\nu\rho}F_{\rho\sigma}F^{\sigma\mu}\biggl],
\label{action-q}
\end{eqnarray}
%\end{widetext}
where
${L}_{GB}:=R^2-4R_{\mu\nu}R^{\mu\nu}+R_{\mu\nu\rho\sigma}R^{\mu\nu\rho\sigma}$
is the Gauss-Bonnet Lagrangian and $\kappa_n:=\sqrt{8\pi G_n}$,
where $G_n$ is $n$-dimensional gravitational constant. The Maxwell
field strength, or the Faraday tensor, is given by
$F_{\mu\nu}:=\partial_\mu A_\nu-\partial_\nu A_\mu$ where $A^\mu$ is
the vector potential. The parameters $\alpha$, $g$, $c_1$, and $c_2$
are real constants.

The action (\ref{action}) with $n=10$ arises in the low-energy limit of
heterotic string theory with constant dilaton. 
Indeed, in the
low-energy limit of the ten-dimensional $E_8 \times E_8$ or $SO(32)$
heterotic string theory with a constant dilaton $\phi_0$ and turning
off the higher form fields, the following Lagrangian is
realized~\cite{gs1987,kmp2007}:
\begin{eqnarray}
\label{low}
\! \! \! \! \! \!  \! \! L_{{\rm low}}&=&\frac{1}{2\kappa_{10}^2}R-\frac{1}{4g^2}F_{\mu\nu}F^{\mu\nu}
+\frac{\alpha'h}{16\kappa_{10}^2}{L}_{GB} 
-\frac{3\alpha'h\kappa_{10}^2}{64}\biggl[(F_{\mu\nu}F^{\mu\nu})^2-
4F_{\mu\nu}F^{\nu\rho}F_{\rho\sigma}F^{\sigma\mu}\biggl],\\
h&:=&e^{-\kappa_{10}\phi_0/\sqrt{2}},
\end{eqnarray}
where the constant $\alpha'$ stands for the inverse string tension.
The above Lagrangian is a particular case of the one considered here
(\ref{action}) with $n=10$, $\Lambda=0$, $\alpha=\alpha'h/8$,
$c_1=-3\alpha'h\kappa_{10}^2/64(<0)$, and $c_2=-4c_1$.

The gravitational equations following from the variation of the
action (\ref{action}) read
\begin{eqnarray}
{\ma G}^\mu_{~~\nu} &:=&{G}^\mu_{~~\nu} +\alpha {H}^\mu_{~~\nu} +
\Lambda \delta^\mu_{~~\nu}=\kappa_n^2T^\mu_{~~\nu}, \label{beq}\\
{G}_{\mu\nu}&:=&R_{\mu\nu}-{1\over 2}g_{\mu\nu}R,\\
{H}_{\mu\nu}&:=&2\Bigl(RR_{\mu\nu}-2R_{\mu\alpha}R^\alpha_{~\nu}-
2R^{\alpha\beta}R_{\mu\alpha\nu\beta}
+R_{\mu}^{~\alpha\beta\gamma}R_{\nu\alpha\beta\gamma}\Bigr)
-{1\over 2}g_{\mu\nu}{L}_{GB}, \label{def-H}
\end{eqnarray}
where the energy-momentum tensor $T_{\mu\nu}$ is given by
\begin{eqnarray}
T_{\mu\nu}&=&\frac{1}{g^2}\biggl(F_{\mu\rho}F_{\nu}^{~\rho}-\frac14 g_{\mu\nu}{\cal F}\biggl)
+2c_1\biggl(\frac12 g_{\mu\nu}{\cal F}^2-4F_{\mu\rho}F_{\nu}^{~\rho}{\cal F}\biggl) \nonumber \\
&+&2c_2\biggl(\frac12 g_{\mu\nu}F_{\lambda\rho}F^{\rho\sigma}F_{\sigma\tau}F^{\tau\lambda}-
4F_{\mu\rho}F^{\rho\sigma}F_{\sigma\tau}F^{\tau}_{~~\nu}\biggl), \label{n-em}\\
{\cal F}&:=&F_{\mu\nu}F^{\mu\nu}.
\end{eqnarray}
The Maxwell equation with the gauge-correction terms reads
\begin{eqnarray}
\nabla_\nu\biggl(-\frac{1}{g^2}F^{\mu\nu}+8c_1{\cal
F}F^{\mu\nu}+8c_2F^{\mu\rho}F_{\rho\sigma}F^{\sigma\nu}\biggl)=0.
\label{max}
\end{eqnarray}

\subsection{Ans{\"a}tze}
Now we consider an Ansatz for the spacetime geometry such that the
$n$-dimensional spacetime $({\ma M}^n, g_{\mu \nu })$ is given by a
warped product of an $(n-2)$-dimensional Einstein space $(K^{n-2},
\gamma _{ij})$ and a two-dimensional orbit spacetime $(M^2, g_{ab})$
under the isometries of $(K^{n-2}, \gamma _{ij})$. Namely, the line
element is given by
\begin{eqnarray}
g_{\mu \nu }d x^\mu d x^\nu =g_{ab}(y)d y^ad y^b +r^2(y) \gamma
_{ij}(z) d z^id z^j , \label{eq:ansatz}
\end{eqnarray}
where $a,b = 0, 1$ while $i,j = 2, ..., n-1$. Here $r$ is a scalar
on $(M^2, g_{ab})$ and
$\gamma_{ij}$ is the metric on $(K^{n-2}, \gamma _{ij})$ with
its sectional curvature $k = \pm 1, 0$.

The $(n-2)$-dimensional Einstein space satisfies
\begin{equation}
\overset{(n-2)}{R}{}_{ijkl}=\overset{(n-2)}{C}{}_{ijkl}+k(\gamma_{ik}\gamma_{jl}-\gamma_{il}\gamma_{jk}),
\end{equation}
where $\overset{(n-2)}{C}{}_{ijkl}$ is the Weyl tensor. The
superscript $(n-2)$ means that the geometrical quantity are defined
on $(K^{n-2}, \gamma _{ij})$. Note that if the Weyl tensor is
identically zero, $(K^{n-2}, \gamma _{ij})$ is a space of constant
curvature. The Riemann tensor is contracted to give
\begin{equation}
\overset{(n-2)}{R}{}_{ij}=k(n-3)\gamma_{ij},\quad \overset{(n-2)}{R}=k(n-2)(n-3).
\end{equation}
In this paper, we consider an Einstein space $(K^{n-2}, \gamma
_{ij})$ satisfying the following condition
\begin{equation}
\overset{(n-2)}{C}{}^{iklm}\overset{(n-2)}{C}{}_{jklm}=\Theta \delta^i_{~~j}, \label{hc}
\end{equation}
where $\Theta$ is constant and non-negative since $(K^{n-2}, \gamma _{ij})$ is an Euclidean space. 
The condition (\ref{hc}) was originally
introduced  by Dotti and Gleiser for the compatibility with the
Einstein-Gauss-Bonnet equations (\ref{beq}) with generic coupling
constants and called the {\it horizon condition}~\cite{dg2005}.
(See~\cite{dot2010} for the classification of the submanifold
depending on the coupling constants.) 
Decomposed geometric tensors of this class of spacetime are presented in Appendix B in~\cite{maeda2010}. 

Nontrivial examples of the Einstein space satisfying the horizon
condition (\ref{hc}) are presented
in~\cite{maeda2010,bcgz2009,md2007}. 
An example of the Einstein space
satisfying the horizon condition that we will consider below is
given by a product space of arbitrary number of
two-dimensional spaces of constant curvature ${K}^2$ with the same
warp factor. In this case, the constant $\Theta$ is given by
$\Theta=2(n-3)(n-4)k^2$. (This is a particular case of the result
shown in Appendix A in~\cite{maeda2010}.)
It is easy to see that if $K^2$ is flat, the
resulting space $(K^{n-2}, \gamma _{ij})$ is nothing but a $(n-2)$-dimensional flat space.
Although the authors do not know concrete non-trivial examples of the Einstein space with non-zero $\Theta$ for some sets of $k$ and $n$ (for $k=0$ with any $n$, for example), we also consider such cases in this paper.

Since ${\cal G}_{ij}$ is proportional to $\gamma_{ij}$, the
energy-momentum tensor must have the following form
\begin{eqnarray}
T_{\mu \nu }d x^\mu d x^\nu = T_{ab}(y)d y^a d y^b+p(y)r^2(y)\gamma
_{ij}d z^id z^j, \label{em}
\end{eqnarray}
where $p(y)$ is a scalar function on $(M^2, g_{ab})$. In analogy
with the spacetime Ansatz (\ref{eq:ansatz}), we look for an
electromagnetic field of the form
\begin{eqnarray}
A_{\mu}dx^\mu=A_a(y)dy^a+A_i(z)dz^i,
\end{eqnarray}
which in turn implies that the Faraday tensor reads
\begin{eqnarray}
F_{\mu \nu }d x^\mu \wedge d x^\nu = F_{ab}(y)d y^a \wedge d
y^b+F_{ij}(z)d z^i \wedge d z^j.
\end{eqnarray}
Here $F_{ab}(y)$ and $F_{ij}(z)$ are identified with the electric
and magnetic components, respectively. For the magnetic component,
we add the following assumption
\begin{eqnarray}
\gamma^{kl}F_{ik}F_{jl}=C^2\gamma_{ij},\label{mag-F}
\end{eqnarray}
where $C$ is a constant. The origin of this condition lies in the
fact that in the case without the gauge-correction terms this
condition is not an input but rather a consequence of the field
equations. 
Hence, it is reasonable to assume Eq.~(\ref{mag-F}) in the presence of the gauge corrections.

%======================================%
%<<<<<<<<<<<< SECTION I  >>>>>>>>>>>>>>%
%======================================%

%%%%%%%%%%%%%%%%%%%%%%%%%%%%%%%%%%%%%%%%%%%%%%
\section{Magnetic solutions}
%%%%%%%%%%%%%%%%%%%%%%%%%%%%%%%%%%%%%%%%%%%%%%
In this section, we obtain magnetic solutions under the assumptions presented in the previous section.
We first determine the possible form of the metric in the next subsection, namely we establish a generalized version of the Jebsen-Birkhoff theorem.
This can be performed without showing the existence of the non-trivial solution of the gauge-corrected Maxwell equation.
The problem of the existence will be studied subsequently.

\subsection{The Jebsen-Birkhoff theorem}
In the vacuum case with $\Theta=0$, the generalized Jebsen-Birkhoff theorem was
shown under the assumption that $(D_a r)(D^a r) \ne 0$
in~\cite{wiltshire1986,birkhoff-gb,birkhoff-gb2}, where $D_a$ is the
covariant derivative on $(M^2, g_{ab})$. For the null case $(D_a
r)(D^a r) = 0$~\cite{mn2008}, on the other hand, there are the
Nariai-Bertotti-Robinson type solutions~\cite{nbr} as in the case
with or without the Maxwell field in general
relativity~\cite{nbr-gr} and in the Einstein-Gauss-Bonnet
gravity~\cite{md2007,nbr-gb,nbr-gb2}. In the case of $\Theta \ne 0$,
the generalized Jebsen-Birkhoff theorem for the vacuum case was
shown in~\cite{maeda2010}.

Here, we only consider the case where $(D_a r)(D^a r) \ne 0$, which
in turn implies that the line element of the spacetime may be
written as
\begin{eqnarray}
ds^2=-g(t,r)e^{-\delta(t,r)}dt^2+\frac{1}{g(t,r)}dr^2+r^2\gamma_{ij}dz^idz^j.
\label{metricAnsatz}
\end{eqnarray}
In this coordinate system, the Maxwell invariant scalar
reads
\begin{eqnarray}
{\cal F}=2F_{tr}F^{tr}+\frac{(n-2)C^2}{r^4},
\end{eqnarray}
while the components of the energy-momentum tensor are given by
%\begin{widetext}
\begin{eqnarray}
T^a_{~~b}&=&\biggl[\frac{1}{2g^2}\biggl(F_{tr}F^{tr}-\frac{(n-2)C^2}{2r^4}\biggl)+2c_1
\biggl(-3F_{tr}F^{tr}+\frac{(n-2)C^2}{2r^4}\biggl)\biggl(2F_{tr}F^{tr}+\frac{(n-2)C^2}{r^4}\biggl)\nonumber \\
&+&2c_2\biggl(-3(F_{tr}F^{tr})^2+\frac{(n-2)C^4}{2r^8}\biggl)\biggl]\delta^a_{~~b}, \\
T^i_{~~j}&=&\biggl[-\frac{1}{2g^2}\biggl(F_{tr}F^{tr}+\frac{(n-6)C^2}{2r^4}\biggl)+2c_1
\biggl(F_{tr}F^{tr}+\frac{(n-10)C^2}{2r^4}\biggl)\biggl(2F_{tr}F^{tr}+\frac{(n-2)C^2}{r^4}\biggl) \nonumber \\
&+&2c_2\biggl((F_{tr}F^{tr})^2+\frac{(n-10)C^4}{2r^8}\biggl)\biggl]\delta^i_{~~j}.
\end{eqnarray}
%\end{widetext}
In the above expressions, we put both the electric component $F_{ab}$ and the magnetic component $F_{ij}$. 
The purely electric case, i.e., $F_{ij}\equiv0$, was fully studied in~\cite{ap2009}. 

Hereafter we consider the purely magnetic case, i.e., $F_{ab}\equiv0$.
Then, the energy-momentum tensor becomes
\begin{eqnarray}
T^a_{~~b}&=&\biggl[-\frac{(n-2)C^2}{4g^2r^4}+\frac{(n-2)\{(n-2)c_1+c_2\}C^4}{r^8}\biggl]\delta^a_{~~b}, \\
T^i_{~~j}&=&\biggl[-\frac{(n-6)C^2}{4g^2r^4}+\frac{(n-10)\{(n-2)c_1+c_2\}C^4}{r^8}\biggl]\delta^i_{~~j}.
\end{eqnarray}
The fact that $T^a_{~~b}\propto \delta^a_{~~b}$ implies ${\ma
G}^{t}_{~~r}={\ma G}^{r}_{~~t}=0$. The integrations of these
constraints restrict the function $g(t,r)$ to be independent of the
variable $t$, i.e. $g(t,r)=f(r)$. Subsequently, the combination
$({\cal G}^t_{~~t}-{\cal
G}^r_{~~r})-\kappa_n^2(T^t_{~~t}-T^r_{~~r})=0$ gives rise to two
different possibilities, namely $\delta(t,r)={\bar \delta}(t)$ or
\begin{eqnarray}
f(r)=k+\frac{r^2}{2(n-3)(n-4)\alpha}. \label{special}
\end{eqnarray}

Let us first consider the latter case. Putting Eq.~(\ref{special})
in the left-hand side of the field equation (\ref{beq}), we obtain
\begin{eqnarray}
{\cal G}^a_{~~b}&=&\biggl[\Lambda+\frac{(n-1)(n-2)}{8\alpha(n-3)(n-4)}-\frac{(n-2)\alpha\Theta}{2r^4}\biggl]\delta^a_{~~b}, \\
{\cal G}^i_{~~j}&=&\biggl[\Lambda+\frac{(n-1)(n-2)}{8\alpha(n-3)(n-4)}-\frac{(n-6)\alpha\Theta}{2r^4}\biggl]\delta^i_{~~j}.
\end{eqnarray}
Hence, for $C\ne 0$, we obtain the following constraints on the
constants of the problem:
\begin{equation}
(n-2)c_1+c_2=0, \quad C^2=\frac{2g^2\alpha\Theta}{\kappa_n^2},\quad
1+\frac{8(n-3)(n-4)\alpha\Lambda}{(n-1)(n-2)}=0 \label{CS}
\end{equation}
with an arbitrary metric function $\delta(t,r)$. It is interesting
to note that for  $C=0$ and $\Theta=0$, this solution reduces to the vacuum solution obtained
in~\cite{birkhoff-gb}.

On the other hand, in the case of $\delta(t,r)={\bar \delta}(t)$, we can set ${\bar
\delta}(t)\equiv 0$ without loss of generality by redefining the
time coordinate. Then, the metric (\ref{eq:ansatz})
reduces to the following simple spacetime with only one unknown
function $f(r)$:
\begin{eqnarray}
ds^2=-f(r)dt^2+\frac{dr^2}{f(r)}+r^2\gamma_{ij}dz^idz^j.
\label{metric2}
\end{eqnarray}
The metric function $f(r)$ is obtained by integrating the gravitational
equations ${\ma G}^{a}_{~~b}=\kappa_n^2T^{a}_{~~b}$ as
%\begin{widetext}
\begin{equation}
f(r)=k+\frac{r^2}{2{\tilde\alpha}}\Biggl(1\mp\sqrt{1+4{\tilde\alpha}{\tilde\Lambda}+\frac{4{\tilde\alpha}{\tilde
M}}{r^{n-1}}+\frac{2{\tilde\alpha}(\kappa_n^2C^2-2g^2{\tilde\alpha}{\tilde\Theta})}
{(n-5)g^2r^{4}}-\frac{8\kappa_n^2{\tilde\alpha}C^4\{(n-2)c_1+c_2\}}{(n-9)r^{8}}}\Biggl),\label{f-GB}
\end{equation}
%\end{widetext}
where ${\tilde M}:=4\kappa_n^2M/[(n-2)V_{n-2}^k]$, ${\tilde
\alpha}:=(n-3)(n-4)\alpha$,
${\tilde\Lambda}:=2\Lambda/[(n-1)(n-2)]$,
${\tilde\Theta}:=\Theta/[(n-3)(n-4)]$, and $M$ is a constant. In the
asymptotically flat vacuum case ($k=1$, $\Theta=0$, $C=0$, and $\Lambda=0$ with the upper sign in (\ref{f-GB})), $M$
gives the Arnowitt-Deser-Misner (ADM) mass.
We emphasize that Eq.~(\ref{CS}) may be satisfied in this solution.
Under the assumptions that $(D_a r)(D^a r)\ne 0$ and the relation
\begin{equation}
1+\frac{8(n-3)(n-4)\alpha\Lambda}{(n-1)(n-2)} \ne 0, \label{noCS}
\end{equation}
the spacetime (\ref{metric2}) with the metric function (\ref{f-GB}) is the unique form of the solution.

Various comments can be made concerning this solution. Firstly,
there are two branches of solutions corresponding to the sign in
front of the square root in Eq.~(\ref{f-GB}), stemming from the
quadratic nature of the field equations. Only the solution with the
upper sign, that we call the GR branch, has a general relativistic
(GR) limit as $\alpha \to 0$ given by
\begin{equation}
f(r)=k-{\tilde\Lambda}r^2-\frac{{\tilde M}}{r^{n-3}}-\frac{\kappa_n^2C^2}{2(n-5)g^2r^{2}}
+\frac{2\kappa_n^2C^4\{(n-2)c_1+c_2\}}{(n-9)r^{6}}.\label{f-GR}
\end{equation}
(In contrast, there is only one branch of real solutions in the third-
order Lovelock gravity~\cite{ds2005}.) Secondly, the metric function (\ref{f-GB}) reduces to the solutions
obtained by Dotti and Gleiser~\cite{dg2005} for $C=0$, by Boulware
and Deser, and independently by Wheeler~\cite{bdw} for $\Theta=0$,
$C=0$, $k=1$, and $\Lambda=0$ and by Lorenz-Petzold and
independently by Cai for $\Theta=0$ and
$C=0$~\cite{lorenz-petzold1988,cai2002}.
Lastly, we see that the metric function (\ref{f-GB}) is not well-defined for $C\ne 0$ with $n=5$ or $n=9$.
However, it is shown in the next subsection that there is no magnetic solution in odd dimensions, namely we have $C \equiv 0$ for odd $n$.

%%%%%%%%%%%%%%%%%%%%%%%%%%%%%%%%%%%%%%%%%%%%%%
\subsection{Non-existence in odd dimensions}
%%%%%%%%%%%%%%%%%%%%%%%%%%%%%%%%%%%%%%%%%%%%%%
We have shown the possible form of the metric for the magnetic solution (\ref{f-GB}).
This does not ensure that there exists a non-trivial magnetic
components of the Faraday tensor satisfying Eqs.~(\ref{max}) and
(\ref{mag-F}).
Under the assumptions presented in the previous section, we prove the non-existence of magnetic
solutions in any odd dimensions.
The proof of this statement is given in~\cite{opz2008} for a more general class of spacetimes but
in order for the paper to be self-contained we present it here in a compact form.
First we obtain $\mbox{det}(F_{ij})\equiv 0$ in odd dimensions by the anti-symmetric nature of $F_{ij}$, explicitly shown by $\mbox{det}(F_{ij})=\mbox{det}(F_{ji})=\mbox{det}(-F_{ij})=(-1)^{n-2}\mbox{det}(F_{ij})$.
Taking the determinant of Eq.~(\ref{mag-F}), we obtain
\begin{eqnarray}
\mbox{det}(F_{il})\mbox{det}(F_{jk})\mbox{det}(\gamma^{lk})=C^{2(n-2)}\mbox{det}(\gamma_{ij}),\label{mag-F2}
\end{eqnarray}
which gives $C\equiv 0$ for odd $n$. Combining the trace of
Eq.~(\ref{mag-F}), which is $F_{ij}F^{ij}=(n-2)C^2$, together with the fact
that $\gamma_{ij}$ is an Euclidean metric, we conclude $F_{ij}\equiv 0$ in any odd dimensions.

%%%%%%%%%%%%%%%%%%%%%%%%%%%%%%%%%%%%%%%%%%%%%%
\subsection{Exact magnetic solutions in even dimensions}
%%%%%%%%%%%%%%%%%%%%%%%%%%%%%%%%%%%%%%%%%%%%%%
Next, we show that there exists a non-trivial magnetic solution in even dimensions.
We first review the monopole-type magnetic solution in
four dimensions. In this case, the Faraday tensor $F_{\mu \nu }d
x^\mu \wedge d x^\nu =Q_{\rm m}{h(\theta)}d \theta \wedge d \phi$ is
the solution for any $k$, where we adopt the coordinates on $(K^{2},
\gamma _{ij})$ such that
$\gamma_{ij}dz^idz^j=d\theta^2+h(\theta)^2d\phi^2$, where
\begin{equation}
h(\theta):=\left\lbrace
\begin{array}{l}
\sin(\theta)~~~~~~~\mbox{for}\quad k=1,\\
1~~~~~~~~~~~~~~\mbox{for}\quad k=0,\\
\sinh(\theta)~~~~~~\mbox{for}\quad k=-1.
\end{array}
\right. \label{h}
\end{equation}
Note that in this case, the constant $C$ is given by $C^2=Q_{\rm
m}^2$.

In higher dimensions, the existence problem of the magnetic solution is highly non-trivial even in general relativity without the
gauge-correction terms except for $k=1$ with $\Theta=0$ where the
magnetic solutions are ruled out. Indeed, for $k=1$ with $\Theta=0$,
the manifold $(K^{n-2}, \gamma_{ij})$ is maximally symmetric with
positive curvature. In the standard Maxwell case, $F_{ij}$ is a
harmonic $2$-form on $(K^{n-2}, \gamma_{ij})$ since it satisfies
$D_{j}F^{ij}=0$, where $D_{i}$ is the covariant derivative on
$(K^{n-2}, \gamma_{ij})$. It is well-known that if the manifold
$(K^{n-2}, \gamma_{ij})$ is compact and its second Betti number is
zero, $F_{ij}\equiv 0$ is satisfied. This is sufficient to prove the
non-existence of magnetic solutions with $k=1$ and
$\Theta=0$ for $n \ge 5$~\cite{nozawa}. In the presence of the
gauge-correction terms, this argument is no longer valid since the
tensor $F_{ij}$ is not necessarily a harmonic $2$-form.

In what follows, we establish the existence of magnetic
solutions with or without the gauge-correction terms for some
special class of Einstein space by extending the standard
four-dimensional monopole result in higher even dimensions. We
consider the Einstein space given as the $(n-2)$-dimensional product
space of $(n-2)/2$ two-dimensional spaces of constant curvature
$K^{n-2} \approx \underset{(n-2)/2}{\underbrace{{K}^2\times
\cdots\times {K}^2}}$ with the same constant warp factor. The metric
on such $(K^{n-2}, \gamma_{ij})$ is given by
\begin{eqnarray}
\gamma_{ij}=&&\frac{1}{n-3}\mbox{diag}({\bar \gamma}_{a_1b_1},
{\bar \gamma}_{a_2b_2},\cdots, {\bar \gamma}_{a_{(n-2)/2}b_{(n-2)/2}}),\\
&&{\bar \gamma}_{a_\sigma b_\sigma}dz^{a_\sigma}dz^{b_\sigma}=
d\theta_\sigma^2+h(\theta_\sigma)^2d\phi_\sigma^2.
\end{eqnarray}
where $\sigma=1,2,\cdots,(n-2)/2$, for which we have $k=\pm1,0$ and
$\Theta=2(n-3)(n-4)k^2$.
As said before,
the manifold $(K^{n-2}, \gamma_{ij})$ is maximally symmetric for $k=0$.
Since a two-dimensional space of non-positive constant curvature can be compactified by certain identifications, the $(n-2)$-dimensional Einstein space $K^{n-2} \approx {K}^2\times
\cdots\times {K}^2$ can be also compactified.

On the above Einstein space, it is easy to show that the following
Faraday tensor satisfies the condition~(\ref{mag-F}):
\begin{eqnarray}
F_{\mu \nu }d x^\mu \wedge d x^\nu =Q_{\rm m}\sum_{\sigma=1}^{(n-2)/2}({h(\theta_\sigma)}d \theta_\sigma \wedge d \phi_\sigma),
\end{eqnarray}
where $Q_{\rm m}^2 \equiv C^2/(n-3)^2$.
This Faraday tensor satisfies the gauge-corrected Maxwell equation~(\ref{max}) as shown below.

The gauge-corrected Maxwell equation~(\ref{max}) can be written as
\begin{equation}
0=\partial_\nu\biggl[\sqrt{-\det(g_{\mu\nu})} 
\biggl(-\frac{1}{g^2}F^{\mu\nu}+8c_1{\cal F}F^{\mu\nu}+8c_2F^{\mu\rho}F_{\rho\sigma}F^{\sigma\nu}\biggl)\biggl].
\end{equation}
For our metric, we obtain
\begin{equation}
-\det(g_{\mu\nu})=\frac{r^{2(n-2)}}{(n-3)^{n-2}}\prod_{\sigma=1}^{(n-2)/2}h(\theta_{\sigma})^2.
\end{equation}
Hence, for $\mu=a$, the gauge-corrected Maxwell equation gives
\begin{equation}
0=\partial_b\biggl[r^{n-2}\biggl(-\frac{1}{g^2}F^{ab}+8c_1{\cal
F}F^{ab}+8c_2F^{ad}F_{df}F^{fb}\biggl)\biggl],
\end{equation}
which is trivially satisfied for the magnetic case.
For $\mu=a_{\sigma}$, we obtain
%\begin{widetext}
\begin{equation}
0=\partial_{b_{\sigma}}\biggl[h(\theta_{\sigma})\biggl(-\frac{1}{g^2}F^{a_{\sigma}b_{\sigma}} 
+8c_1{\cal F}F^{a_{\sigma}b_{\sigma}}+8c_2F^{a_{\sigma}d_{\sigma}}F_{d_{\sigma}f_{\sigma}}F^{f_{\sigma}b_{\sigma}}\biggl)\biggl].\label{max-sub}
\end{equation}
%\end{widetext}
Using the following expressions:
\begin{eqnarray}
F_{\theta_{\sigma}\phi_{\sigma}}&=&Q_{\rm m}h,\quad F^{\theta_{\sigma}\phi_{\sigma}}=\frac{(n-3)^2}{r^4h}Q_{\rm m},\\
{\cal F}&=&\frac{(n-2)(n-3)^2}{r^4}Q_{\rm m}^2,
\end{eqnarray}
we show that inside the bracket in Eq.~(\ref{max-sub}) is independent from $z^{b_{\sigma}}$.
Hence, Eq.~(\ref{max-sub}) is also satisfied.

We finally close this section by briefly commenting about the
existence of the dyonic solution, that is the solution with both
electric and magnetic charges. Although it is difficult to obtain
the explicit form of the metric function in the dyonic case with the
gauge corrections, this task is render possible in the absence of these
terms. The solution in this case is given by
%\begin{widetext}
\begin{equation}
f(r)=k+\frac{r^2}{2{\tilde\alpha}}\Biggl(1\mp\sqrt{1+4{\tilde\alpha}{\tilde\Lambda}
+\frac{4{\tilde\alpha}{\tilde M}}{r^{n-1}}-\frac{4\kappa_n^2
{\tilde\alpha} Q_{\rm e}^{2}}{(n-2)(n-3)g^2
r^{2(n-2)}}+\frac{2{\tilde\alpha}(\kappa_n^2C^2-2g^2{\tilde\alpha}{\tilde\Theta})}{(n-5)g^2r^{4}}}\Biggl),
\end{equation}
%\end{widetext}
and the non-zero electric component of the Faraday tensor is
\begin{equation}
F_{tr}=\frac{Q_{\rm e}}{r^{n-2}}, \label{ftr}
\end{equation}
where $Q_{\rm e}$ is a constant corresponding to the electric charge.

%======================================%
%<<<<<<<<<<<< SECTION I  >>>>>>>>>>>>>>%
%======================================%

%%%%%%%%%%%%%%%%%%%%%%%%%%%%%%%%%%%%%%%%%%%%%%%%%%%%%%%%%%%%%%%%%
\section{Properties of the magnetic black holes with gauge corrections}
%%%%%%%%%%%%%%%%%%%%%%%%%%%%%%%%%%%%%%%%%%%%%%%%%%%%%%%%%%%%%%%%%
In this section, we analyze the properties of the magnetic solution
(\ref{metric2}) with Eq.~(\ref{f-GB}). We first point out that the
coupling constants of the gauge-correction terms $c_1$ and $c_2$
appear in the metric function only through the combination
$(n-2)c_1+c_2$. This shows a sharp contrast with the electric case, in which they appear in the following more rigid form
$2c_1+c_2$~\cite{ap2009}. It is also appealing to
note that in the purely magnetic case, the power of the
potential of the Maxwell term as well as the gauge-correction term
appearing in the metric is independent of the number of dimensions.
This is clearly in contrast with the solution in the purely
electric case. We note that, in the monopole type solution with the Yang-Mills field, 
the power of the matter term in the metric function is also constant for 
$n \ge 6$~\cite{ym}.

%%%%%%%%%%%%%%%%%%%%%%%%%%%%%%%%%%%%%
\subsection{Curvature singularities}
%%%%%%%%%%%%%%%%%%%%%%%%%%%%%%%%%%%%%

In the spacetime given by (\ref{f-GB}), there are at most two
classes of curvature singularities. There is a curvature singularity
localized at the center $r=0$ while the other is at $r=r_{\rm b}$,
where $r_{\rm b}$ corresponds to the possible zero of the
square-root piece of the metric function (\ref{f-GB}).
Both at $r=0$ and $r=r_{\rm b}$, the Kretschmann invariant
\begin{eqnarray}
K&:=&R_{\mu\nu\rho\sigma}R^{\mu\nu\rho\sigma} \nonumber \\
&=&\biggl(\frac{d^2f}{dr^2}\biggl)^2+\frac{2(n-2)}{r^2}\biggl(\frac{df}{dr}\biggl)^2+\frac{2(n-2)(n-3)}{r^4}(k-f)^2
\end{eqnarray}
blows up.
The latter is called the branch singularity since two branches of solutions meet there.
The branch singularity is a characteristic singularity in higher-curvature gravity located at a finite physical radius in general.
As a direct consequence of the existence of branch
singularity is that the domain of the radial coordinate $r$ can not
be extended from $0$ to $\infty$. The appearance of the branch
singularity sharply depends on the parameters of the solution, and
the location $r=r_{\rm b}$ is given by solving the following
algebraic equation $B(r_{\rm b})=0$, where
\begin{equation}
B(r_{\rm b}):=1+4{\tilde\alpha}{\tilde\Lambda}+\frac{4{\tilde\alpha}{\tilde
M}}{r_{\rm
b}^{n-1}}+\frac{2{\tilde\alpha}(\kappa_n^2C^2-2g^2{\tilde\alpha}{\tilde\Theta})}{(n-5)g^2r_{\rm
b}^{4}} 
-\frac{8\kappa_n^2{\tilde\alpha}C^4\{(n-2)c_1+c_2\}}{(n-9)r_{\rm b}^{8}}.\label{h}
\end{equation}
The physical domain of the radial coordinate $r$ is given by $B(r)>0$.

Here we only consider the physically reasonable situations in which
$\alpha>0$, $1+4{\tilde\alpha}{\tilde\Lambda} \ge 0$, and $M \ge 0$
are satisfied. The first condition is imposed by string theory,
while the second inequality ensures the existence of the maximally
symmetric solution. The last condition means that the parameter
$M$ which is assimilated to the mass is positive. With this respect,
it is a non-trivial issue to see whether the parameter $M$ can be
identified as the mass of a black hole in the present case because
the spacetime has a non-trivial boundary. However, in the vacuum
case, $M$ coincides with the well-defined quasi-local mass and
satisfies the first law of the black-hole thermodynamics together
with the Wald entropy~\cite{maeda2010}. For these reasons, we call
$M$ the mass parameter. Finally, it is simple to see that under
the condition $(n-9)[(n-2)c_1+c_2] \le 0$ with a sufficiently large
value of the magnetic constant $C^2$, the solution is free from branch
singularities.

%%%%%%%%%%%%%%%%%%%%%%%%%%%%%%%%%%
\subsection{Asymptotic structure}
%%%%%%%%%%%%%%%%%%%%%%%%%%%%%%%%%%

Let us next consider the asymptotic structure of our solution, that
is the behavior for $r \to \infty$. For $\Theta=0$, this
spacetime is at least locally asymptotically flat or
(anti-)de~Sitter ((A)dS) for $\lambda=0$ or $\lambda(<)>0$,
respectively, in the sense that
\begin{eqnarray}
R_{~~~\sigma \rho }^{\mu \nu }|_{r\to \infty}&=&\lambda\left( \delta _{\sigma }^{\mu }\delta _{\rho }^{\nu
}-\delta_{\rho }^{\mu }\delta _{\sigma }^{\nu }\right),\\
\lambda &:=&-\frac{1}{2{\tilde\alpha}}\left(1\mp \sqrt{1+4{\tilde\alpha}{\tilde\Lambda}}\right).
\end{eqnarray}
In four dimensions, in which $\Theta=0$, the magnetic term
respects the fall-off conditions to the asymptotically flat or AdS
regions. Then, for $k=1$, $M$ corresponds to the
Arnowitt-Deser-Misner (ADM) mass and to the Abbott-Deser (AD) mass
in the asymptotically flat and AdS cases, respectively. In
higher dimensions, on the other hand, the fall-off rate of the
magnetic and the Weyl terms in the metric function is slower than
the mass term. 

The contribution of the higher-order gauge corrections decays more rapidly for $r \to \infty$ than the Maxwell term. 
On the other hand, the gauge-correction term
dominates around the center $r \to 0$ in the generic case and its
contribution is quite sensitive to the sign of $(n-2)c_1+c_2$. 
In the general relativistic case (\ref{f-GR}), unlike in four
dimensions, the magnetic term contributes as the attractive force in
higher dimensions while the higher-order gauge corrections give the
repulsive (attractive) force depending on the sign of the constant
$[(n-2)c_1+c_2]/(n-9)$. As a result, the global structure of the
spacetime can be quite different from the standard
Reissner-Nordstr\"{o}m case.  Finally, we close this section by
stressing that through a fine-tuning between the parameters such as
$(n-2)c_1+c_2=0$, the gauge corrections do not appear in the metric.
In addition, if the magnetic constant has a very precise value $C^2=2g^2{\tilde\alpha}{\tilde\Theta}/\kappa_n^2$, the metric function (\ref{f-GB}) is the same as the generalized Boulware-Deser-Wheeler solution.

%%%%%%%%%%%%%%%%%%%%%%%%%%%%%%%
\subsection{Energy conditions}
%%%%%%%%%%%%%%%%%%%%%%%%%%%%%%%

In fact, the sign of $(n-2)c_1+c_2$ is closely related to the energy
condition. The energy-momentum tensor of our matter field has the
diagonal form as $T^\mu_{~~\nu}=\mbox{diag}(-\mu,p_{\rm r},p_{\rm
t},p_{\rm t},\cdots)$. The physical interpretations of $\mu$,
$p_{\rm r}$ and $p_{\rm t}$ are the energy density, radial pressure
and tangential pressure, respectively. The weak energy condition
(WEC) implies $\mu \ge 0$, $p_{\rm r}+\mu \ge 0$, and $p_{\rm t}+\mu
\ge 0$, while the dominant energy condition (DEC) implies $\mu \ge
0$, $-\mu \le p_{\rm r} \le \mu$, and $-\mu \le p_{\rm t} \le \mu$.
The null energy condition (NEC) implies $p_{\rm r}+\mu \ge 0$, and
$p_{\rm t}+\mu \ge 0$~\cite{he,carroll}. Note that DEC implies WEC
and WEC implies NEC.

For
our matter field, the corresponding energy density, radial pressure,
and the tangential pressure are
\begin{eqnarray*}
\mu&=&\frac{(n-2)C^2}{4g^2r^4}-\frac{(n-2)C^4\{(n-2)c_1+c_2\}}{r^8}, \label{mu} \\
p_{\rm r}&=&-\frac{(n-2)C^2}{4g^2r^4}+\frac{(n-2)C^4\{(n-2)c_1+c_2\}}{r^8}, \label{pr} \\
p_{\rm t}&=&-\frac{(n-6)C^2}{4g^2r^4}+\frac{(n-10)C^4\{(n-2)c_1+c_2\}}{r^8}, \label{pt}
\end{eqnarray*}
from which we obtain
\begin{eqnarray*}
\mu+p_{\rm r}&=&0, \\
\mu-p_{\rm r}&=&\frac{(n-2)C^2}{2g^2r^4}-\frac{2(n-2)C^4\{(n-2)c_1+c_2\}}{r^8}, \\
\mu+p_{\rm t}&=&\frac{C^2}{g^2r^4}-\frac{8C^4\{(n-2)c_1+c_2\}}{r^8}, \\
\mu-p_{\rm t}&=&\frac{(n-4)C^2}{2g^2r^4}-\frac{2(n-6)C^4\{(n-2)c_1+c_2\}}{r^8}.
\end{eqnarray*}
Hence, it is clear that if $(n-2)c_1+c_2 \le 0$, the DEC is
satisfied for any positive $r$ while if $(n-2)c_1+c_2>0$, the NEC is
violated near $r=0$.

%%%%%%%%%%%%%%%%%%%%%%%%%%%%%%%%%%%%%%%
\subsection{Black hole configurations}
%%%%%%%%%%%%%%%%%%%%%%%%%%%%%%%%%%%%%%%

Now we clarify the parameter region where the solution represents a
black hole. A Killing horizon is given by $r=r_{\rm h}$ such that
$f(r_{\rm h})=0$. An outer Killing horizon is defined by $f(r_{\rm
h})=0$ with $df/dr(r_{\rm h})>0$. On the other hand, an inner and
degenerate Killing horizons are characterized by $df/dr(r_{\rm
h})<0$ and $df/dr(r_{\rm h})=0$, respectively. A black hole is
defined by an event horizon, which is an outermost outer Killing
horizon if there exists null infinity. Notice that an outermost
degenerate Killing horizon with $d^2f/dr^2(r_{\rm h})>0$ may also be
an event horizon.

For this purpose, the ${\tilde M}$-$r_{\rm h}$ diagram is quite
useful. (See~\cite{tm2005} for the analysis with or without the
Maxwell electric charge in the case of the maximally symmetric horizon.) 
The ${\tilde M}$-$r_{\rm h}$ relation is
obtained from the equation $f(r_{\rm h})=0$ as
%\begin{widetext}
\begin{eqnarray}
{\tilde M}&=&-{\tilde\Lambda}r_{\rm h}^{n-1}+kr_{\rm h}^{n-3}-
\frac{[\kappa_n^2C^2-2g^2{\tilde\alpha}\{(n-5)k^2+{\tilde\Theta}\}]}
{2(n-5)g^2}r_{\rm
h}^{n-5}+\frac{2\kappa_n^2C^4\{(n-2)c_1+c_2\}}{n-9} r_{\rm
h}^{n-9}, \nonumber \\
&=:&{\tilde M}_{\rm h}(r_{\rm h}).
\end{eqnarray}
On the other hand, the ${\tilde M}$-$r_{\rm b}$ relation is obtained from $B(r_{\rm b})=0$ as
\begin{eqnarray}
{\tilde M}&=&-\frac{1}{4{\tilde\alpha}}(1+4{\tilde\alpha}{\tilde\Lambda})
r_{\rm b}^{n-1}-\frac{(\kappa_n^2C^2-2g^2{\tilde\alpha}{\tilde\Theta})}
{2(n-5)g^2}r_{\rm b}^{n-5}+\frac{2\kappa_n^2C^4\{(n-2)c_1+c_2\}}{(n-9)}
r_{\rm b}^{n-9}, \nonumber \\
&=:&{\tilde M}_{\rm b}(r_{\rm b}).
\end{eqnarray}
%\end{widetext}
We calculate
\begin{equation}
{\tilde M}_{\rm h}(r)-{\tilde M}_{\rm b}(r)=\frac{r^{n-5}(r^2+2{\tilde\alpha}k)^2}{4{\tilde\alpha}},
\end{equation}
and hence ${\tilde M}_{\rm h}\ge {\tilde M}_{\rm b}$ is satisfied
for $\alpha>0$  with equality holding at $r=0$ as well as for
$r^2=-2{\tilde\alpha}k$ as long as ${\tilde\alpha}k<0$.

The number of horizons and the existence of the branch singularity for the given mass $M$ are totally understood by the functional forms of ${\tilde M}_{\rm h}(r)$ and ${\tilde M}_{\rm b}(r)$, respectively.
However, the shape of the two curves ${\tilde M}={\tilde M}_{\rm h}(r)$ and
${\tilde M}={\tilde M}_{\rm b}(r)$ depends on the parameters in a
complicated manner and it is almost hopeless to provide a complete
classification. However, since this work is motivated by the
low-energy action of string theory (\ref{low}), we focus our attention on
the case with $\alpha \ge 0$, $\Lambda=0$, $(n-2)c_1+c_2 \le 0$ and
with $k=1$ (and hence $\Theta=2(n-3)(n-4)$). In this case, the
previous expressions reduce to
\begin{eqnarray}
{\tilde M}_{\rm h}(r)&=&r^{n-3}-\frac{q^2-2(n-3){\tilde\alpha}}
{2(n-5)}r^{n-5}-\frac{2q^4d^2}{n-9}r^{n-9},\label{Mh} \\
{\tilde M}_{\rm b}(r)&=&-\frac{1}{4{\tilde\alpha}}r^{n-1}-
\frac{q^2-4{\tilde\alpha}}{2(n-5)}r^{n-5}-\frac{2q^4d^2}{n-9}r^{n-9},\label{Mb} \\
q^2&:=&\frac{\kappa_n^2C^2}{g^2},\qquad d^2:=-\frac{g^4\{(n-2)c_1+c_2\}}{\kappa_n^2}.
\end{eqnarray}
For later purpose, let us compute the first and second derivatives:
\begin{eqnarray}
\frac{d{\tilde M}_{\rm h}}{dr}&=&(n-3)r^{n-4}-\frac{q^2-2(n-3){\tilde\alpha}}
{2}r^{n-6}-2q^4d^2r^{n-10}, \label{dMh} \\
\frac{d^2{\tilde M}_{\rm h}}{dr^2}&=&(n-3)(n-4)r^{n-5}-\frac{(n-6)[q^2-2(n-3){\tilde\alpha}]}{2}r^{n-7} 
-2(n-10)q^4d^2r^{n-11}, \label{ddMh} \\
%%%%%%%%%%%%%%%%%%%%%%%%%%%%%%%%%%%%%%%%%%%%%%%%%%%%%%%%
\frac{d{\tilde M}_{\rm b}}{dr}&=&-\frac{n-1}{4{\tilde\alpha}}r^{n-2}-
\frac{q^2-4{\tilde\alpha}}{2}r^{n-6}-2q^4d^2r^{n-10}, \label{dMb} \\
\frac{d^2{\tilde M}_{\rm b}}{dr^2}&=&-\frac{(n-1)(n-2)}{4{\tilde\alpha}}r^{n-3}-
\frac{(n-6)[q^2-4{\tilde\alpha}]}{2}r^{n-7}
-2(n-10)q^4d^2r^{n-11}. \label{ddMb}
\end{eqnarray}

We will see that the existence or absence of the horizon depends
on the parameters. The solution with horizons belongs the GR branch
because $f(r)>0$ is satisfied and there is no horizon in the non-GR
branch for $\alpha > 0$ with $k=1$. For $\alpha>0$ with $k=1$, the
branch singularity is in the untrapped region defined by $f(r)>0$.
The asymptotic region $r\to \infty$ is in the untrapped region since
$\lim_{r\to \infty}f(r)=1$ is satisfied for $k=1$ and $\Lambda=0$ in
the GR branch.

In order to clarify the effects of the higher-order correction terms, we will study four cases separately in the following subsections.
There we adopt the unit such that ${\tilde\alpha}=1$ for $\alpha>0$.
For $(n-2)c_1+c_2\ne 0$, we adopt the unit in addition such that $d^2=1$.

\subsubsection{General relativity without gauge corrections}
\begin{figure}[htbp]
\begin{center}
%\rotatebox{-90}{
%\includegraphics[width=1.0\linewidth]{GRc=0.eps}
\subfigure[]{\includegraphics[width=0.45\linewidth]{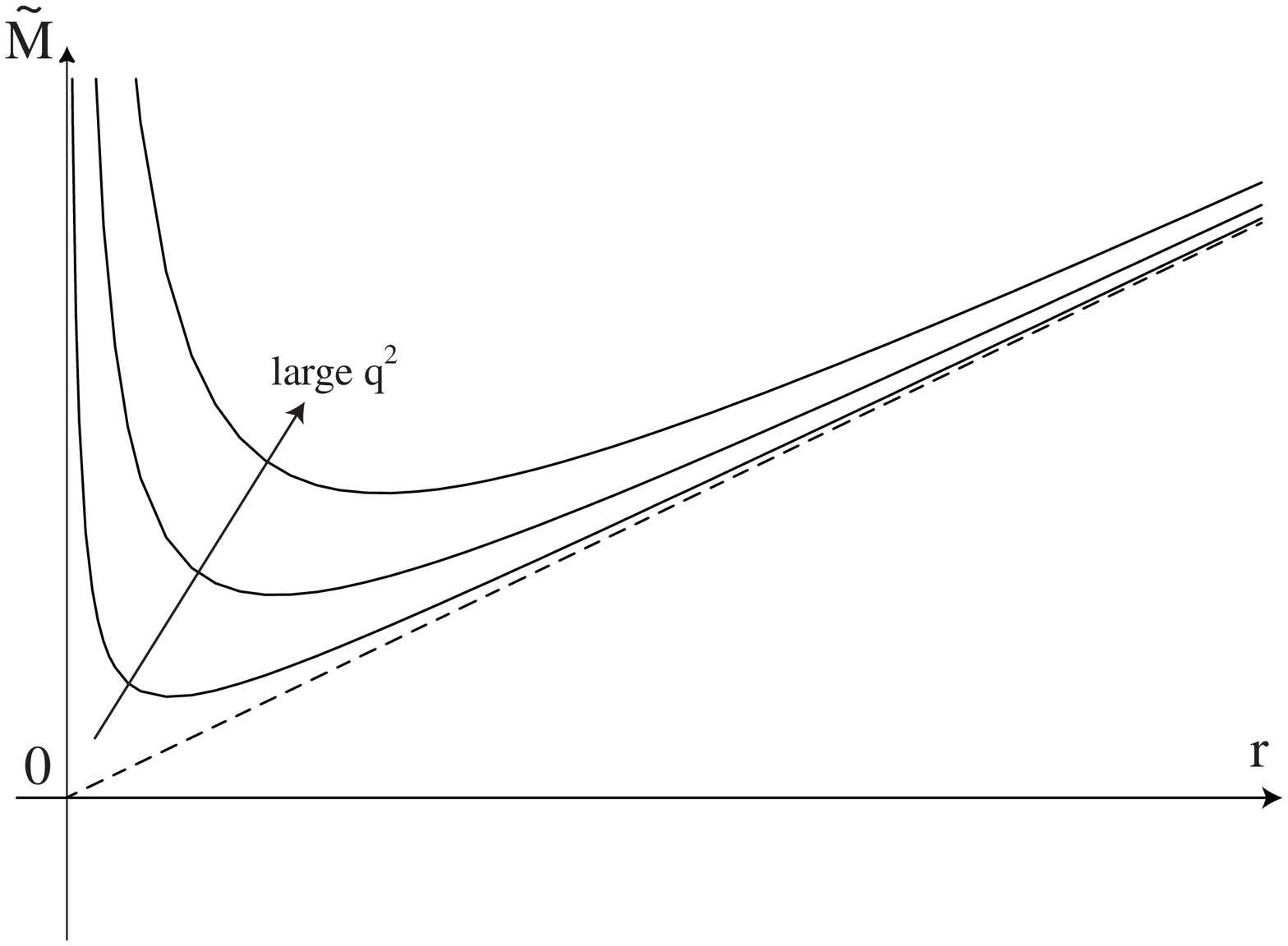}}
\subfigure[]{\includegraphics[width=0.45\linewidth]{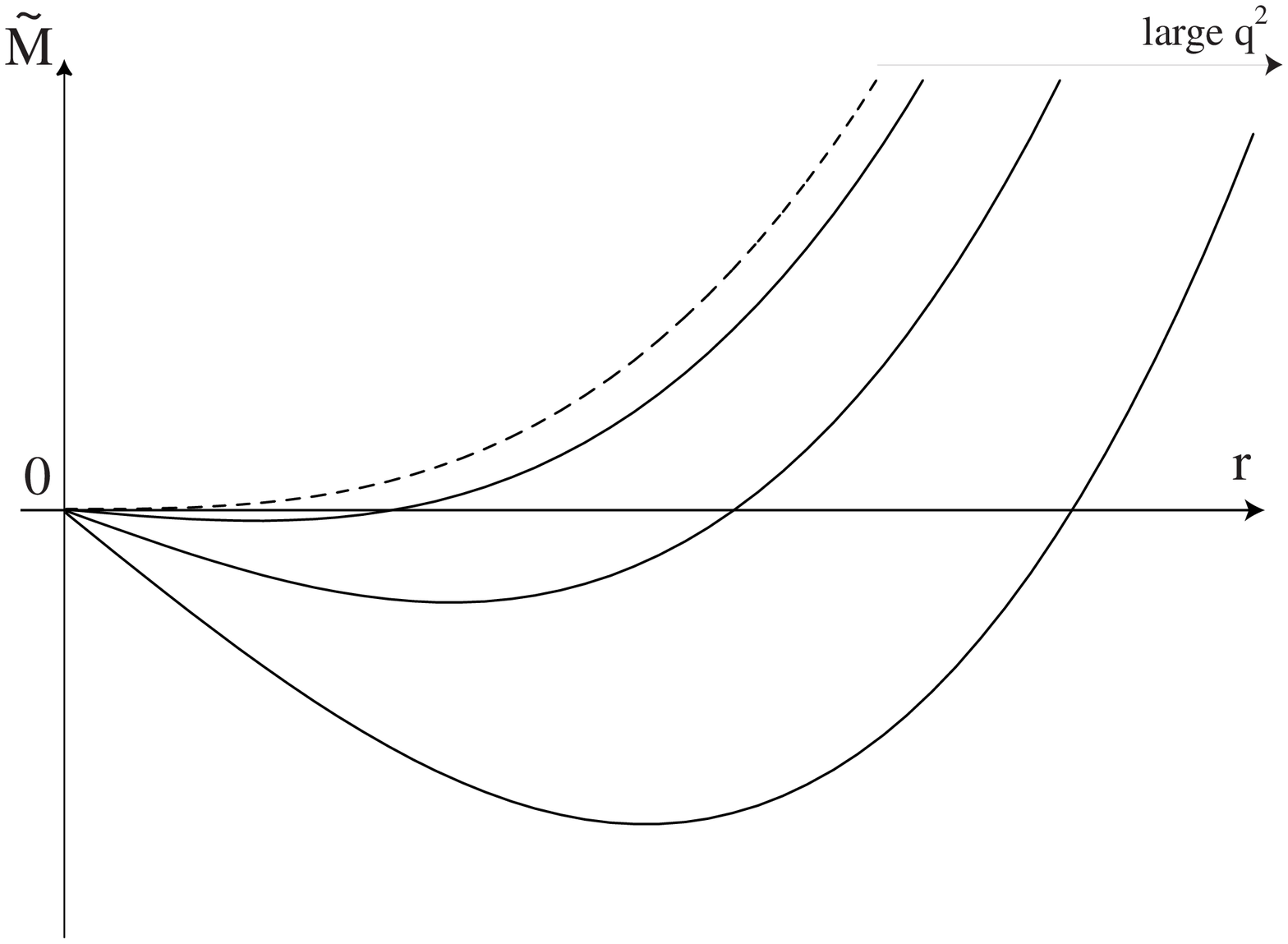}}
%}
\caption{\label{fig1} The function ${\tilde M}={\tilde M}_{\rm
h}(r)$ in the positive-curvature case without a cosmological
constant and gauge corrections in general relativity ($k=1$,
$\Lambda=0$, $\alpha=0$, and $(n-2)c_1+c_2=0$). The parameter
dependence on $q^2$ is shown for (a) $n=4$ and (b) $n=6$. A dashed
curve corresponds to the case with $q^2=0$. The graph for $n\ge 8$
is qualitatively the same as $n=6$.}
\end{center}
\end{figure}

First we consider the simplest case, namely the general relativistic case without gauge corrections.
The ${\tilde M}$-$r_{\rm h}$ diagram given by Eq.~(\ref{Mh}) with $\alpha=d=0$ is qualitatively different between $n=4$ and $n\ge 6$.
Also, it is different between $q^2=0$ and $q^2 \ne 0$ for each $n$.
(See Fig.~\ref{fig1}.)

%-----------<fig>---------------------------
%--------------<fig>-----------------------
For $n=4$, the situation is the same as the Schwarzschild or the
Reissner-Nordstr\"om solution. For $n \ge 6$ with $q^2=0$, there is
one outer horizon for ${\tilde M}>0$, while there is no horizon for
${\tilde M} \le 0$. For $n \ge 6$ with $q^2>0$, ${\tilde M}={\tilde
M}_{\rm h}(r)$ has one local minimum at ${\tilde M}={\tilde M}_{\rm
ex}(<0)$. There is one outer horizon for ${\tilde M} \ge 0$, one
outer and one inner horizons for $0>{\tilde M}>{\tilde M}_{\rm ex}$,
one degenerate horizon for ${\tilde M}={\tilde M}_{\rm ex}$, and no
horizon for ${\tilde M}<{\tilde M}_{\rm ex}$.

\subsubsection{General relativity with gauge corrections}
\begin{figure*}[htbp]
\begin{center}
%\rotatebox{-90}{
%\includegraphics[width=1.0\linewidth]{GRc=0.eps}
\subfigure[]{\includegraphics[width=0.4\linewidth]{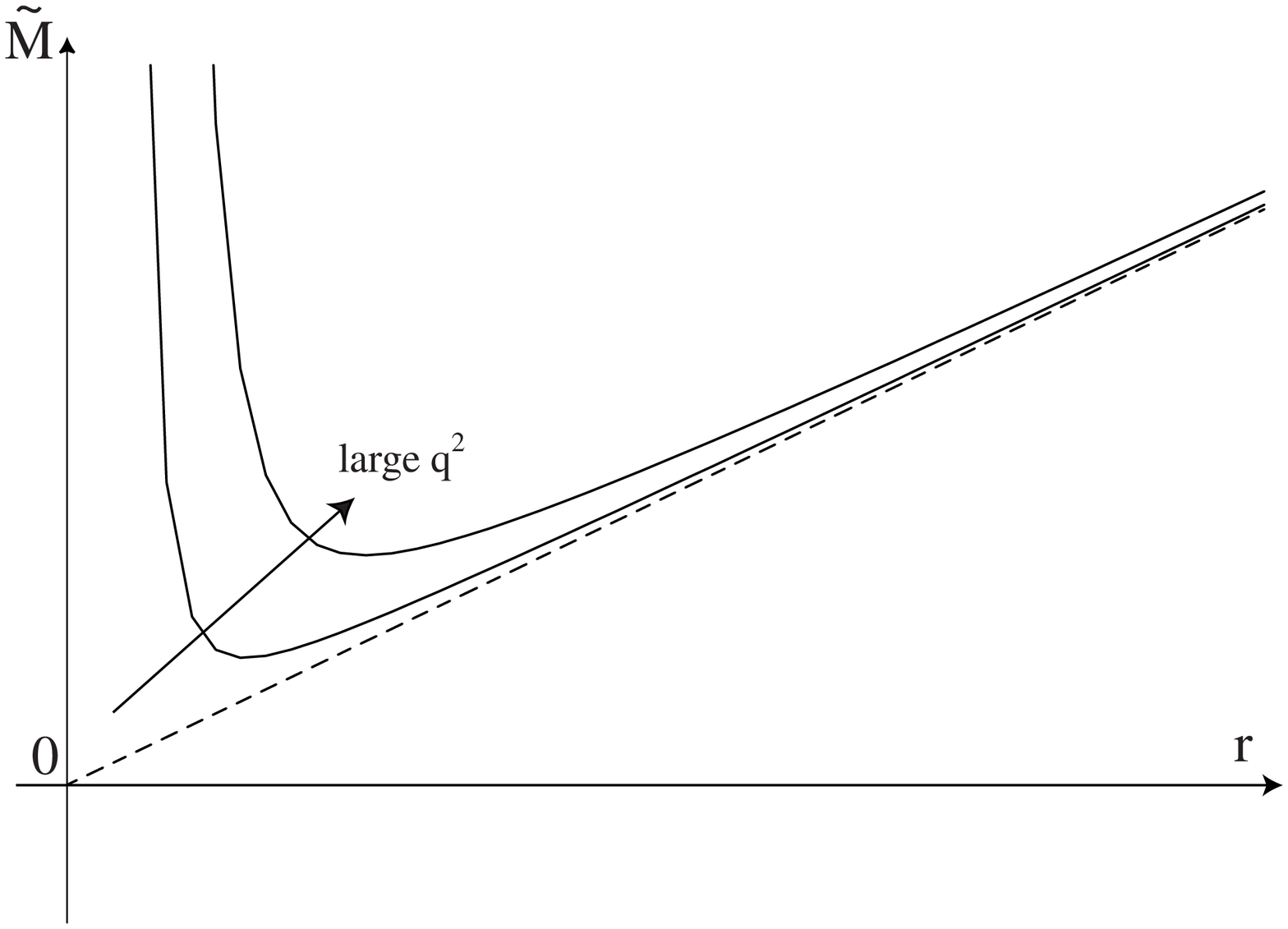}}
\subfigure[]{\includegraphics[width=0.4\linewidth]{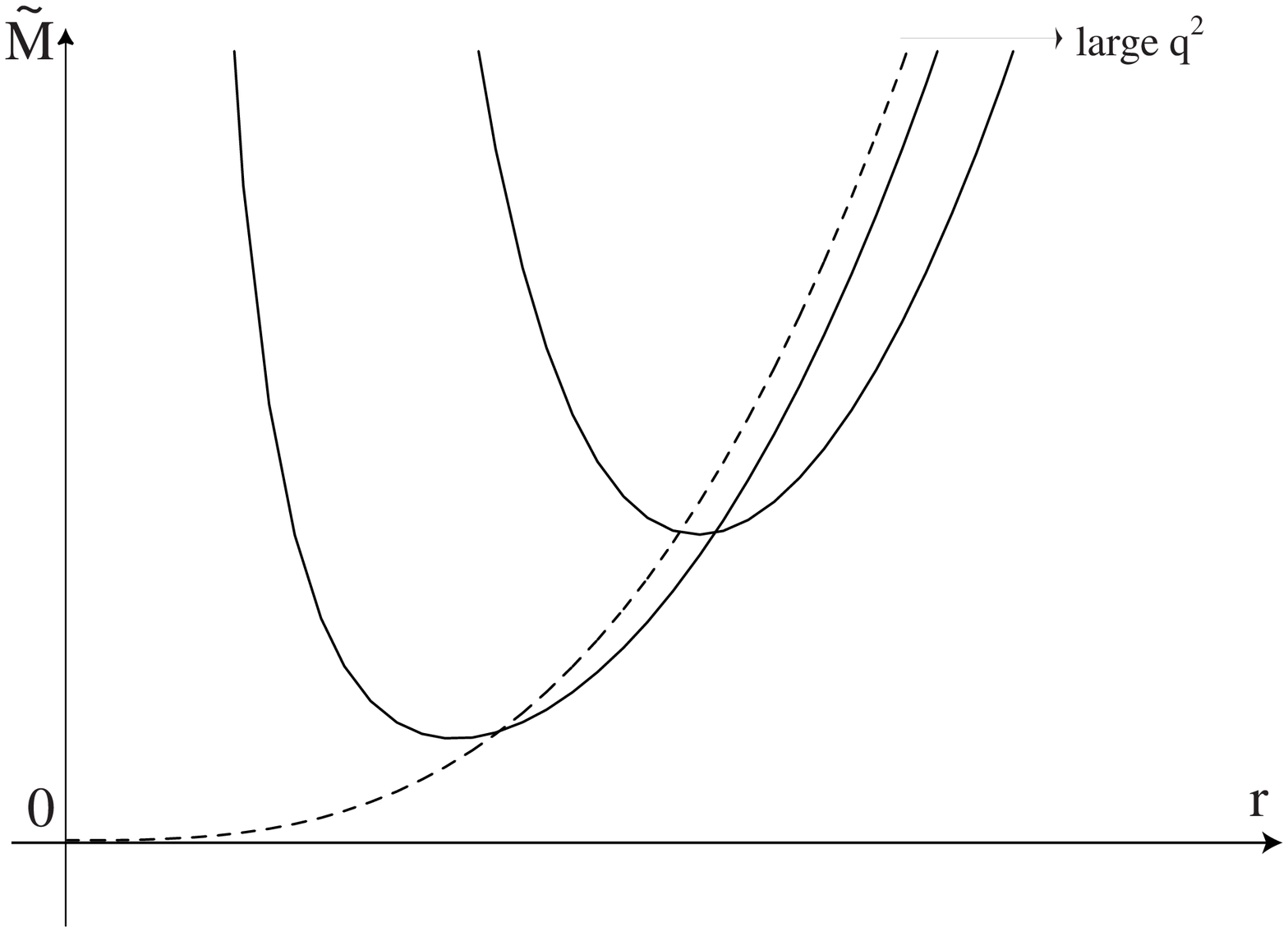}}
\subfigure[]{\includegraphics[width=0.4\linewidth]{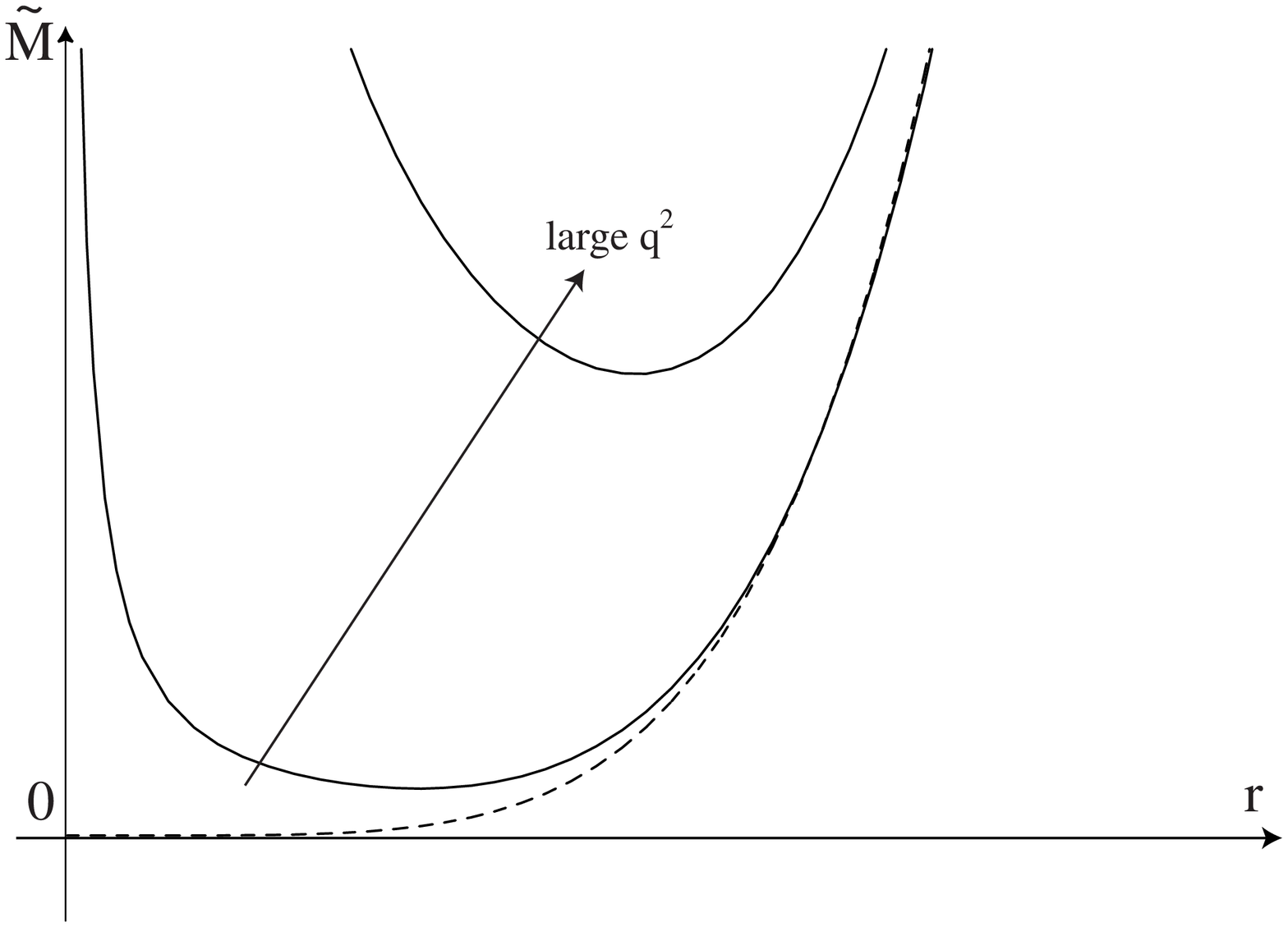}}
\subfigure[]{\includegraphics[width=0.4\linewidth]{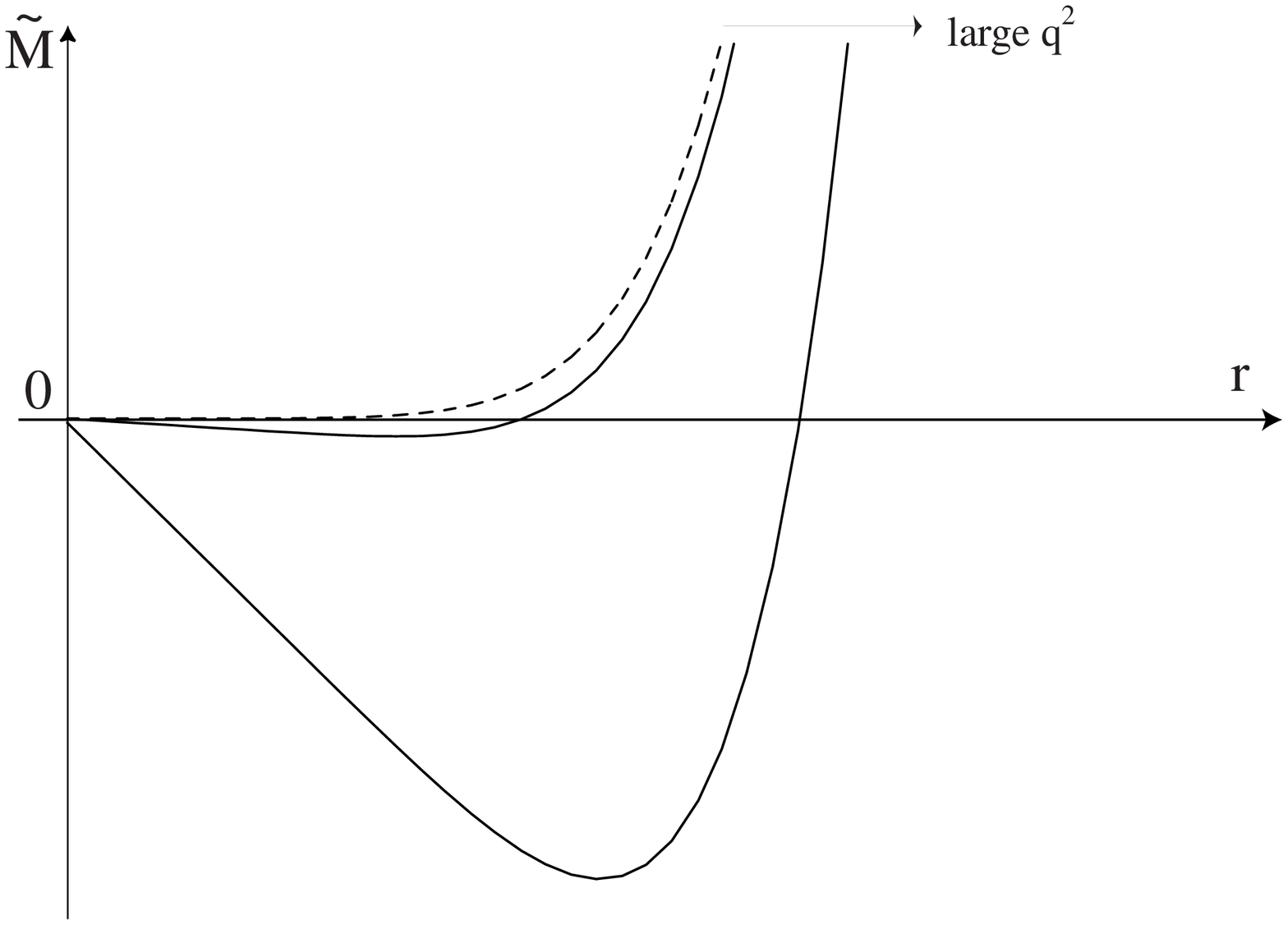}}
%}
\caption{\label{fig2} The function ${\tilde M}={\tilde M}_{\rm
h}(r)$ in the positive-curvature case without a cosmological
constant in general relativity but with gauge corrections ($k=1$, $\Lambda=0$, $\alpha=0$, and
$(n-2)c_1+c_2<0$). The parameter dependence on $q^2$ is shown for (a) $n=4$, (b) $n=6$, (c) $n=8$, and (d) $n=10$. A dashed curve
corresponds to the case with $q^2=0$. The graph for $n\ge 12$ is
qualitatively the same as $n=10$.}
\end{center}
\end{figure*}

Next we consider the effect of the gauge-correction terms in general
relativity. The ${\tilde M}$-$r_{\rm h}$ diagrams given by
Eq.~(\ref{Mh}) with $\alpha=0$ and $d^2=1$ are shown in
Fig.~\ref{fig2}.
%-----------<fig>---------------------------
%--------------<fig>-----------------------
For $n=4$, the situation is qualitatively the same as the case
without gauge corrections. For $n\ge 6$ with $q^2=0$, there is one
outer horizon for ${\tilde M}>0$, while there is no horizon for
${\tilde M} \le 0$. For $n=6,8$ with $q^2>0$, ${\tilde M}={\tilde
M}_{\rm h}(r)$ has one local minimum at ${\tilde M}={\tilde M}_{\rm
ex}(>0)$. There are one outer and one inner horizons for ${\tilde
M}>{\tilde M}_{\rm ex}$, one degenerate horizon for ${\tilde
M}={\tilde M}_{\rm ex}$, and no horizon for ${\tilde M}<{\tilde
M}_{\rm ex}$. For $n \ge 10$ with $q^2>0$, ${\tilde M}={\tilde
M}_{\rm h}(r)$ has one local minimum at ${\tilde M}={\tilde M}_{\rm
ex}(<0)$. There is one outer horizon for ${\tilde M} \ge 0$, one
outer and one inner horizons for $0>{\tilde M}>{\tilde M}_{\rm ex}$,
one degenerate horizon for ${\tilde M}={\tilde M}_{\rm ex}$, and no
horizon for ${\tilde M}<{\tilde M}_{\rm ex}$.

\subsubsection{Einstein-Gauss-Bonnet gravity without gauge corrections}
Now we consider the effect of the Gauss-Bonnet term for $n\ge 6$.
The most drastic change is the existence of the branch singularity.
In the presence of the branch singularity for given $M$, we only
consider the domain of $r$ connecting to the asymptotic region,
namely $r_{\rm b}<r<\infty$.

\begin{figure*}[htbp]
\begin{center}
%\rotatebox{-90}{
%\includegraphics[width=1.0\linewidth]{GRc=0.eps}
\subfigure[]{\includegraphics[width=0.4\linewidth]{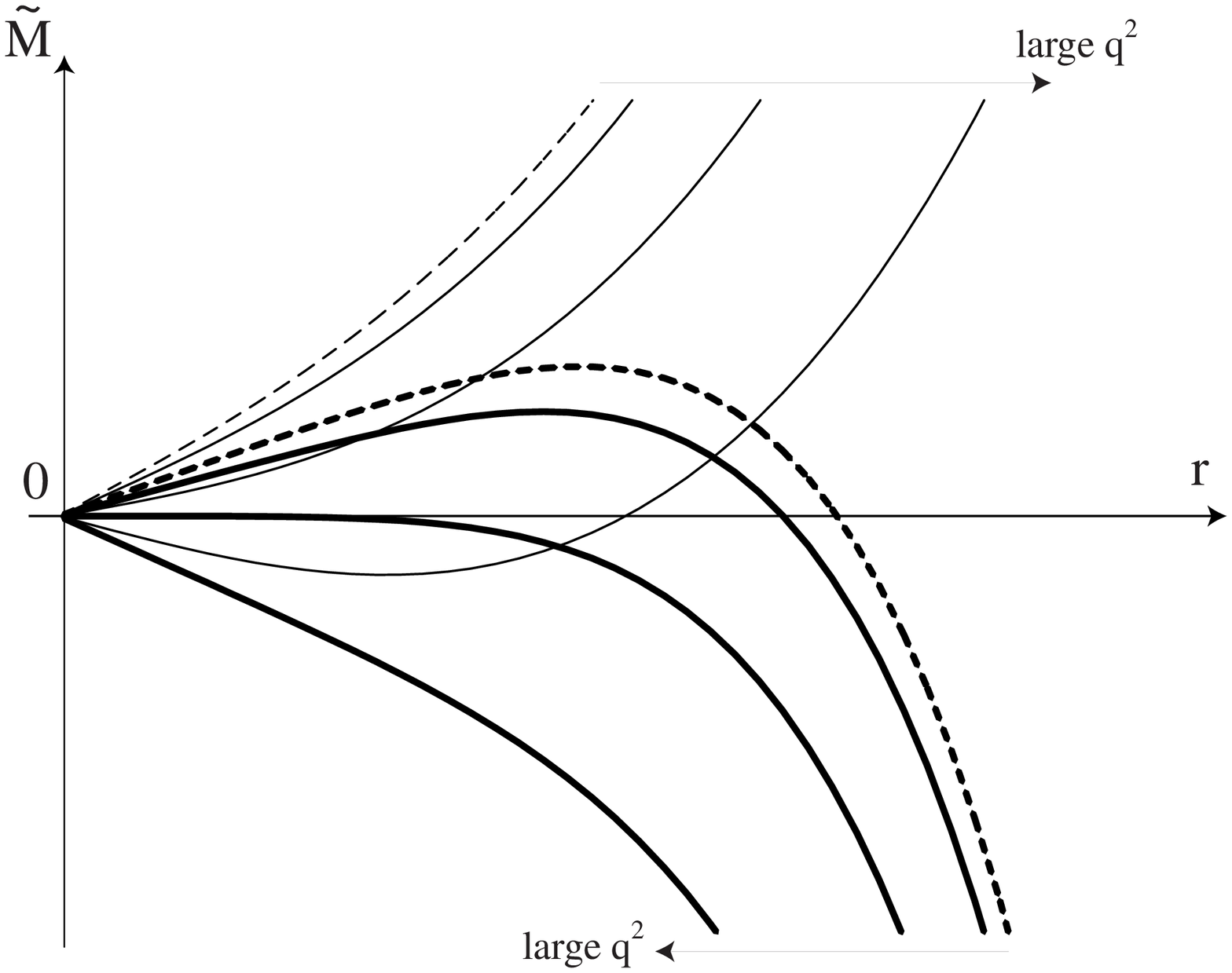}}
\subfigure[]{\includegraphics[width=0.4\linewidth]{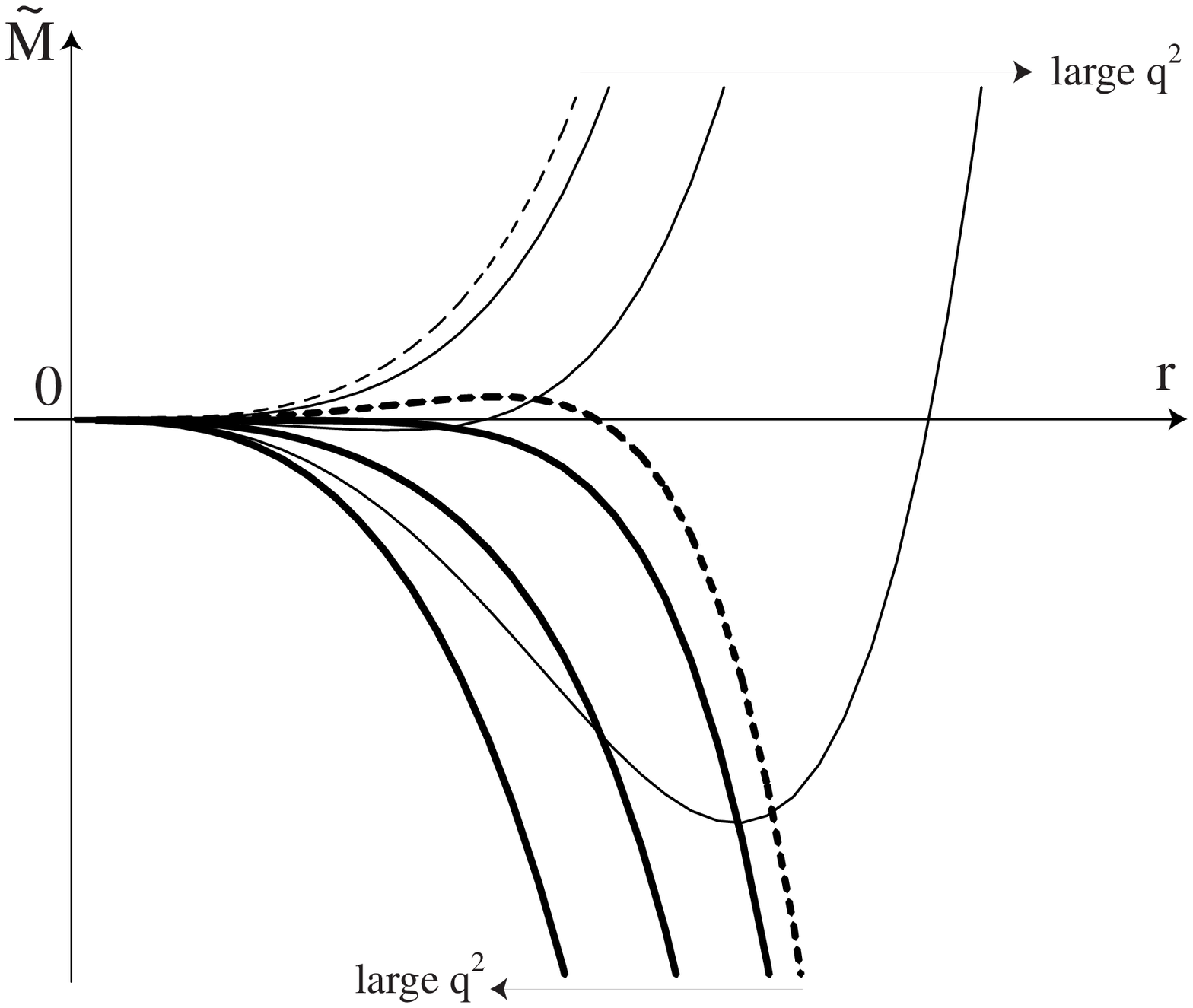}}
\subfigure[]{\includegraphics[width=0.4\linewidth]{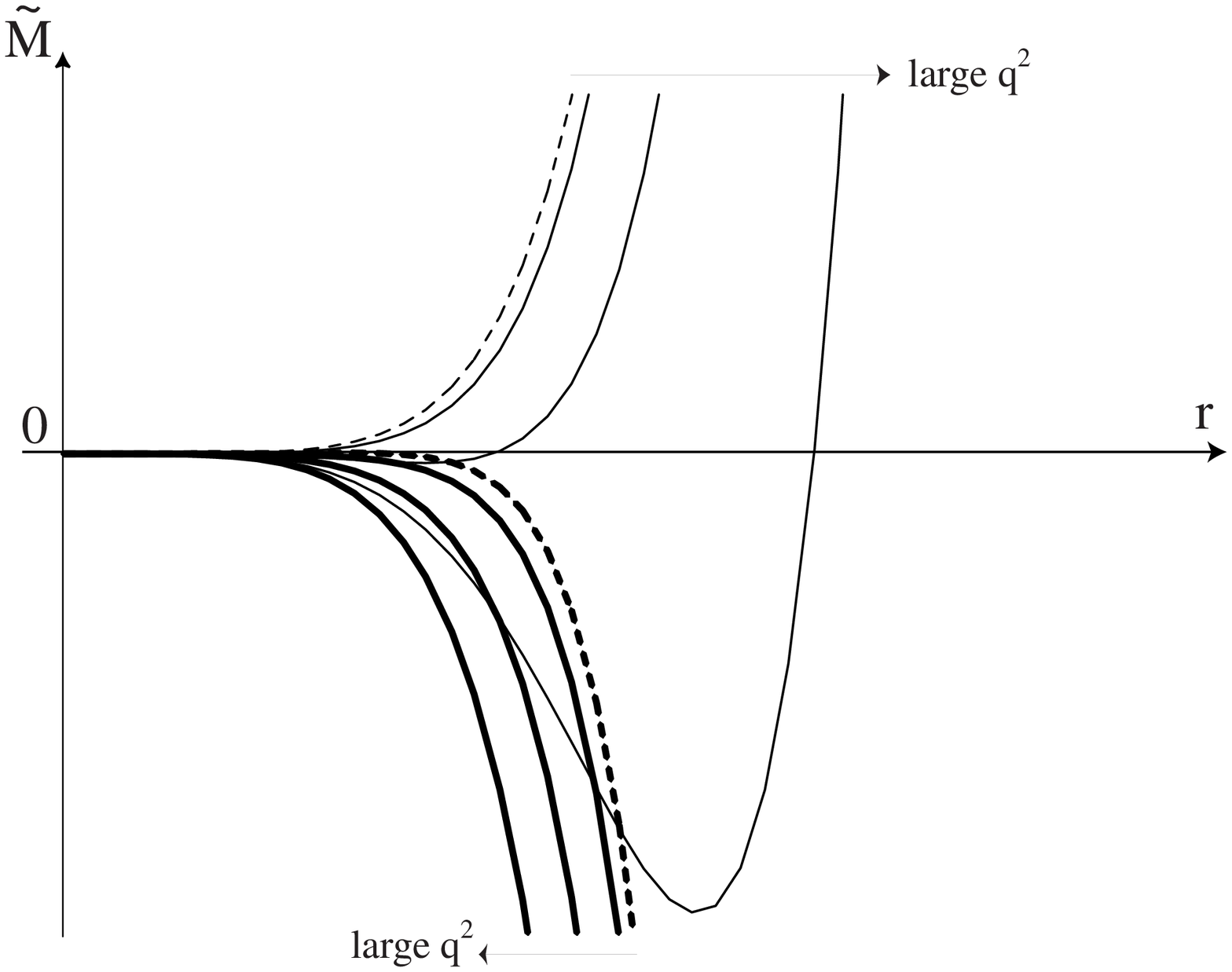}}
%}
\caption{\label{fig3} The functions ${\tilde M}={\tilde M}_{\rm
h}(r)$ and ${\tilde M}={\tilde M}_{\rm b}(r)$ in the positive-curvature case without a cosmological constant and gauge corrections
in Einstein-Gauss-Bonnet gravity ($k=1$, $\alpha>0$, $\Lambda=0$,
and $(n-2)c_1+c_2=0$). The parameter dependence on $q^2$ is shown
for (a) $n=6$, (b) $n=8$, and (c) $n=10$. A thin and a thick curves
correspond to ${\tilde M}={\tilde M}_{\rm h}(r)$ and ${\tilde
M}={\tilde M}_{\rm b}(r)$, respectively. The dashed curves
correspond to the case with $q^2=0$. The physical domain of $r$ is
${\tilde M}>{\tilde M}_{\rm b}$ and ${\tilde M}={\tilde M}_{\rm b}$
is in the untrapped region. The graph for $n\ge 12$ is qualitatively
the same as $n=10$.}
\end{center}
\end{figure*}

We first consider the case without gauge corrections. The ${\tilde
M}$-$r_{\rm h}$ and ${\tilde M}$-$r_{\rm b}$ diagrams, given
respectively by Eqs.~(\ref{Mh}) and (\ref{Mb}) with
${\tilde\alpha}=1$ and $d^2=0$, are shown in Fig.~\ref{fig3}.
%-----------<fig>---------------------------
%--------------<fig>-----------------------
The graphs are qualitatively the same for any $n\ge 6$.
For $0 \le q^2 \le 2(n-3)$, ${\tilde M}={\tilde M}_{\rm h}(r)$ is monotonically increasing from ${\tilde M}=0$.
For $q^2 > 2(n-3)$, ${\tilde M}={\tilde M}_{\rm h}(r)$ has one local minimum at ${\tilde M}={\tilde M}_{\rm ex(h)}(<0)$
For $0 \le q^2 < 4$, ${\tilde M}={\tilde M}_{\rm b}(r)$ has one local maximum at ${\tilde M}={\tilde M}_{\rm ex(b)}(>0)$.
For $q^2 \ge  4$, ${\tilde M}={\tilde M}_{\rm b}(r)$ is monotonically decreasing from ${\tilde M}=0$.

Hence, for $0 \le q^2 < 4$, there is one outer
horizon for ${\tilde M}>{\tilde M}_{\rm ex(b)}$ and no horizon for
${\tilde M}\le {\tilde M}_{\rm ex(b)}$.
For $4 \le q^2 \le 2(n-3)$,
there is one outer horizon for ${\tilde M}>0$ and no horizon for
${\tilde M} \le 0$.
For $q^2>2(n-3)$, there is one outer horizon for ${\tilde M} \ge 0$, one
outer and one inner horizons for $0>{\tilde M}>{\tilde M}_{\rm ex(h)}$,
one degenerate horizon for ${\tilde M}={\tilde M}_{\rm ex(h)}$, and no
horizon for ${\tilde M}<{\tilde M}_{\rm ex(h)}$.

\subsubsection{Einstein-Gauss-Bonnet gravity with gauge corrections}
We finally consider the case where both the Gauss-Bonnet and
gauge-correction terms are present. The ${\tilde M}$-$r_{\rm h}$ and
${\tilde M}$-$r_{\rm b}$ diagrams are given respectively by
Eqs.~(\ref{Mh}) and (\ref{Mb}) with ${\tilde\alpha}=1$ and $d^2=1$
and the parameter dependence is rather complicated.

We first analyze
the behavior of ${\tilde M}={\tilde M}_{\rm b}$ with the help of its
derivative Eq.~(\ref{dMb}).
It is simple to see that an extremum exists if
\begin{eqnarray}
(q^2-4)^2-8(n-1)q^4>0.
\end{eqnarray}
However, the left-hand side of the above inequality can not be
positive for $n \ge 2$ and hence there is no extremum and ${\tilde
M}={\tilde M}_{\rm b}$ is monotonic. Also, it is seen that
$\lim_{r\to \infty}{\tilde M}_{\rm b}(r)=-\infty$ and $\lim_{r\to
\infty}{\tilde M}_{\rm h}(r)=+\infty$. Near $r=0$, we obtain
$\lim_{r\to 0}{\tilde M}_{\rm b}(r)=+\infty$, $\lim_{r\to 0}{\tilde
M}_{\rm h}(r)=+\infty$ for $n \le 8$, while $\lim_{r\to 0}{\tilde
M}_{\rm b}(r)=0$, $\lim_{r\to 0}{\tilde M}_{\rm h}(r)=0$ for $n \ge
10$.

The behavior of ${\tilde M}={\tilde M}_{\rm h}$ can be better
analyzed by its derivatives (\ref{dMh}) and (\ref{ddMh}).
Since the algebraic equation $d{\tilde M}_{\rm h}/dr=0$ is
essentially cubic for $r^2$, it is difficult to provide a rigorous
argument about the behavior of ${\tilde M}={\tilde M}_{\rm h}$.
However, the numerical plots of ${\tilde M}={\tilde M}_{\rm h}$
indicate that there is only one local minimum at ${\tilde M}={\tilde
M}_{\rm ex}$, where ${\tilde M}_{\rm ex}>0$ and ${\tilde M}_{\rm
ex}<0$ are satisfied for $n=6,8$ and $n\ge 10$, respectively.

\begin{figure*}[htbp]
\begin{center}
%\rotatebox{-90}{
%\includegraphics[width=1.0\linewidth]{GRc=0.eps}
\subfigure[]{\includegraphics[width=0.4\linewidth]{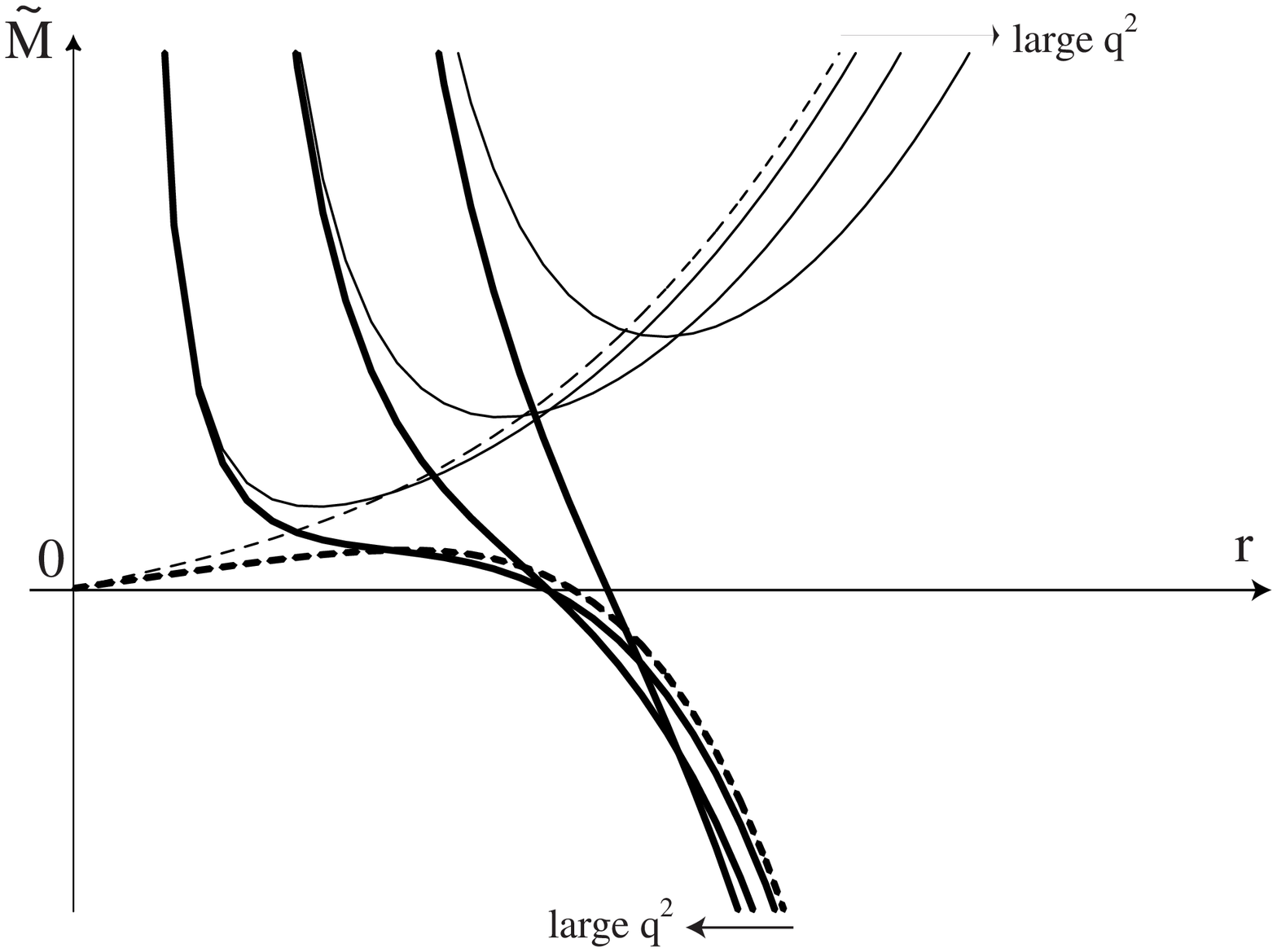}}
\subfigure[]{\includegraphics[width=0.4\linewidth]{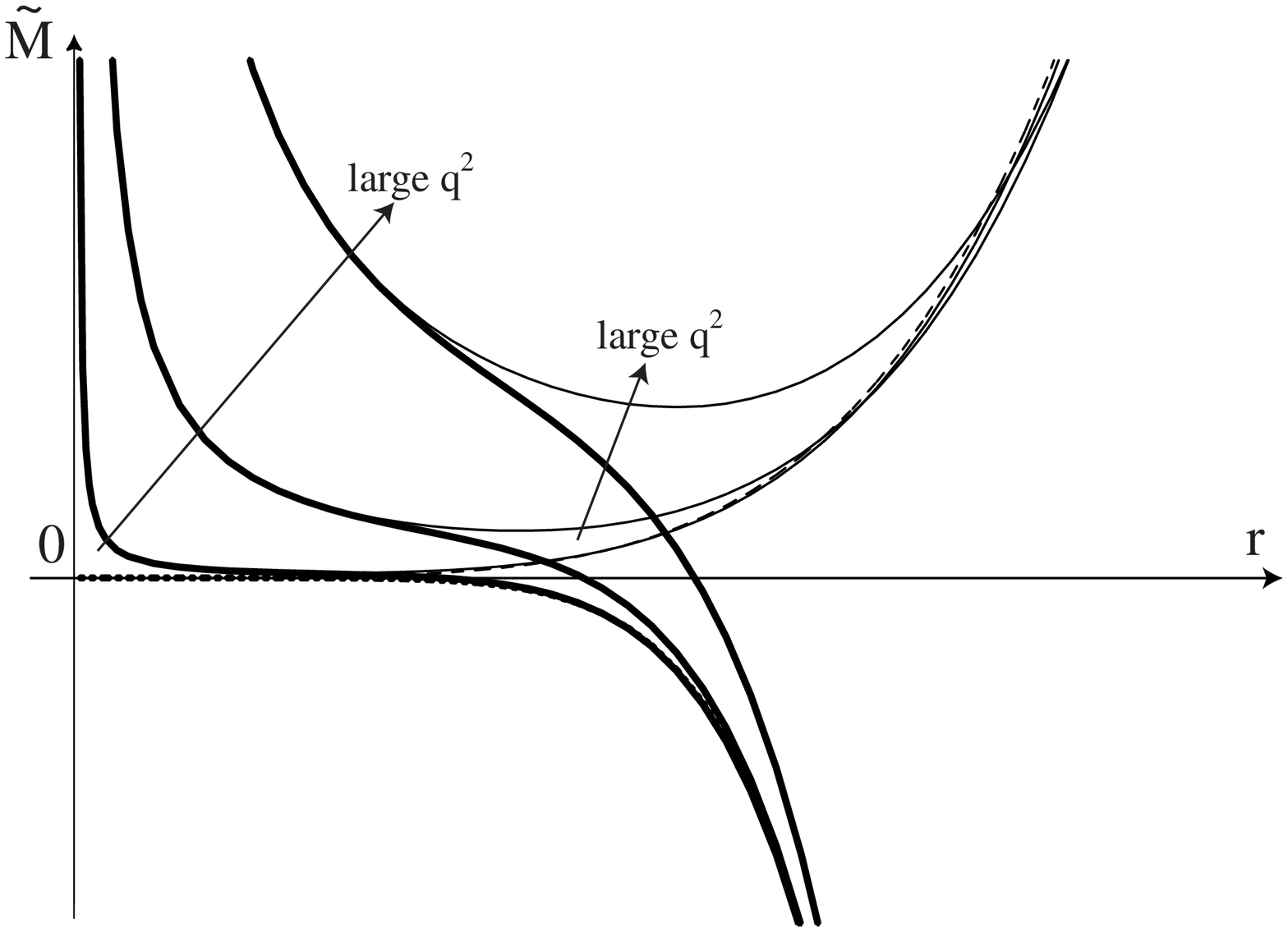}}
\subfigure[]{\includegraphics[width=0.4\linewidth]{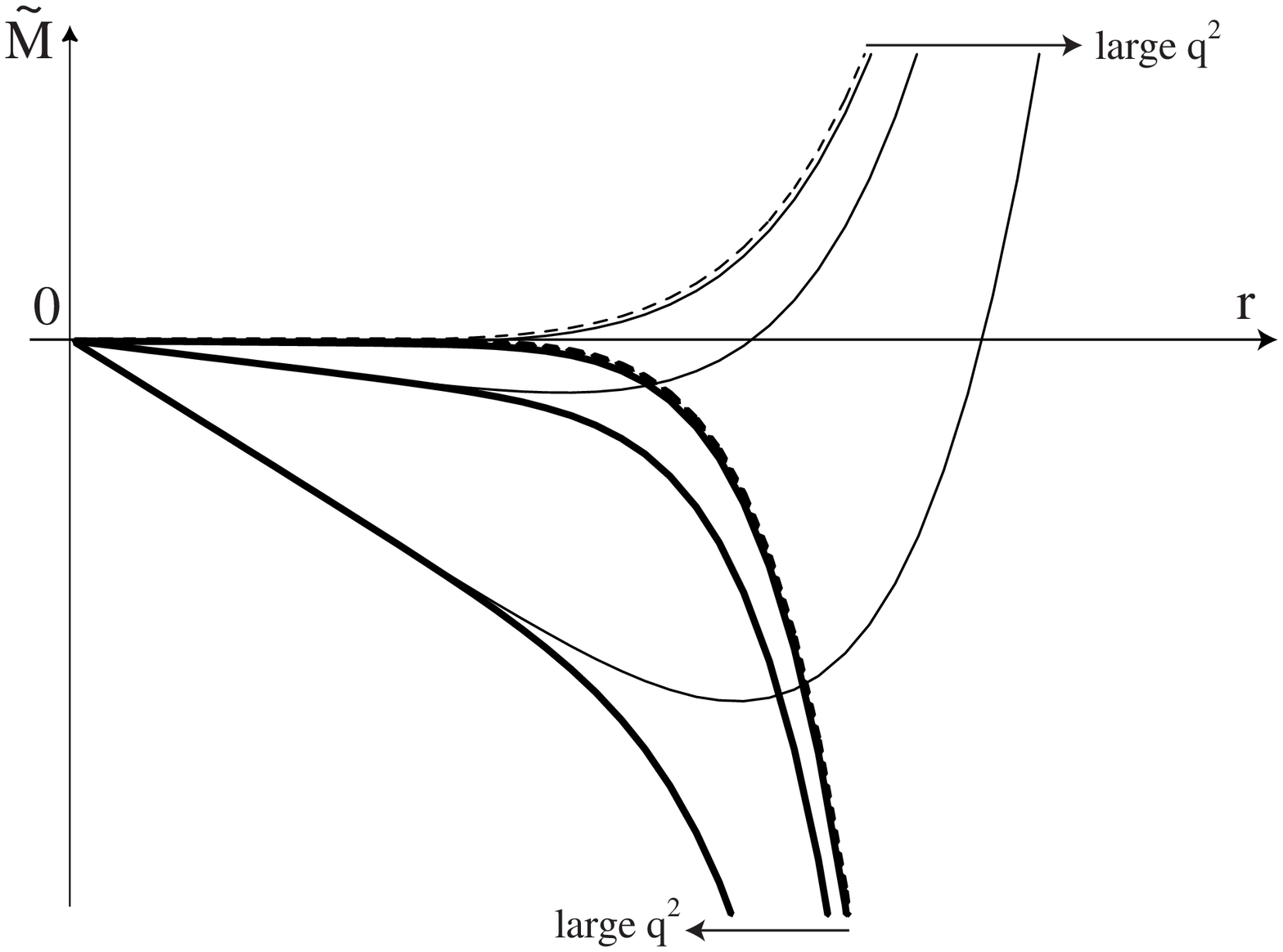}}
%}
\caption{\label{fig4} The functions ${\tilde M}={\tilde M}_{\rm
h}(r)$ and ${\tilde M}={\tilde M}_{\rm b}(r)$ in the positive-curvature case without a cosmological constant in
Einstein-Gauss-Bonnet gravity ($k=1$, $\alpha>0$, $\Lambda=0$, and
$(n-2)c_1+c_2<0$). The parameter dependence on $q^2$ is shown for
(a) $n=6$, (b) $n=8$, and (c) $n=10$. A thin and a thick curves
correspond to ${\tilde M}={\tilde M}_{\rm h}(r)$ and ${\tilde
M}={\tilde M}_{\rm b}(r)$, respectively. The dashed curves
correspond to the case with $q^2=0$. The physical domain of $r$ is
${\tilde M}>{\tilde M}_{\rm b}$ and ${\tilde M}={\tilde M}_{\rm b}$
is in the untrapped region. The graph for $n\ge 12$ is qualitatively
the same as $n=10$.}
\end{center}
\end{figure*}

The ${\tilde M}$-$r_{\rm h}$ and ${\tilde M}$-$r_{\rm b}$ diagrams are shown in Fig.~\ref{fig4}.
%-----------<fig>---------------------------
%--------------<fig>-----------------------
In the case of $q^2=0$, the graphs are the same as the case without
gauge corrections (Fig.~\ref{fig3} with $q^2=0$). For $n=6,8$, there
are one outer and one inner horizons for ${\tilde M}>{\tilde M}_{\rm
ex}$, one degenerate horizon for ${\tilde M}={\tilde M}_{\rm ex}$,
and no horizon for ${\tilde M}<{\tilde M}_{\rm ex}$. For any
${\tilde M}$, there exists a branch singularity. For $n=10$, there
is one outer horizon for ${\tilde M} \ge 0$, one outer and one inner
horizons for $0>{\tilde M}>{\tilde M}_{\rm ex}$, one degenerate
horizon for ${\tilde M}={\tilde M}_{\rm ex}$, and no horizon for
${\tilde M}<{\tilde M}_{\rm ex}$. The branch singularity exists for
${\tilde M}<0$.

%======================================%
%<<<<<<<<<<<< SECTION I  >>>>>>>>>>>>>>%
%======================================%
\section{Summary and discussions}
In the present paper, we have considered the $n(\ge 4)$-dimensional
Einstein-Gauss-Bonnet equations in presence of a cosmological
constant with a matter source given by the Maxwell action with the $F^4$ gauge-correction terms build
up with the Faraday tensor.
This action without a cosmological
constant is realized in the low-energy limit of a class of string
theories. We have assumed that the spacetime geometry is given by a
warped product ${\ma M}^{2} \times {\ma K}^{n-2}$, where ${\ma
K}^{n-2}$ is a $(n-2)$-dimensional Einstein space satisfying a
specific condition (\ref{hc}) and the orbit of the warp factor on
${\ma K}^{n-2}$ is non-null.

Under a few reasonable assumptions, we have established the
generalized Jebsen-Birkhoff theorem for the magnetic
solution which fixes the metric function in a unique form. Using a
simple geometric argument, we have established the non-existence of
such magnetic solutions in any odd dimensions. In even
dimensions, we have obtained magnetic solutions in the
case where ${\ma K}^{n-2}$ is a product manifold of $(n-2)/2$ two-dimensional maximally symmetric spaces with the same
constant warp factors.

The coupling constants of the gauge-correction terms appear in the
metric function in the form of $(n-2)c_1+c_2$ and the
gauge-correction term converges to zero rapidly for $r\to \infty$,
while it dominates in the short distance for $n \le 8$. We have clarified whether
the solution represents a black hole or not depending on the
parameters in the case of $k=1$, $\Lambda=0$, $\alpha\ge 0$,
$(n-2)c_1+c_2 \le 0$, which is the most important case in direct
relation with the string viewpoints. We have established that the
existence of black hole configurations is not only tied to the
existence of the gauge-correction terms, but also to the number of
spacetime dimensions. In the presence of the gauge-correction terms,
the qualitative properties of the magnetic black hole is
rather different if the even dimension $n \le 8$ or if $n\ge 10$.
This is not only because the power of the gauge-correction term in the metric function (\ref{f-GB}) becomes smaller than the mass term for $n\ge 10$, but also because the sign of the gauge-correction term is different for $n \le 8$ and $n\ge 10$.

As a future task, the black-hole thermodynamics of our magnetic
black hole is important. In Einstein-Gauss-Bonnet gravity, this
subject have been intensively investigated with or without the
Maxwell electric charge in the case where ${\ma K}^{n-2}$ is
maximally symmetric. The effect of the Weyl term on the
thermodynamical stability has been recently analyzed for the
Dotti-Gleiser vacuum black hole by one of the
authors~\cite{maeda2010}. However, the thermodynamical aspect of
magnetic black holes in higher dimensions has not been
studied yet even in the standard Maxwell case in general relativity.

Another interesting problem would be to introduce a non-trivial
dilaton since it naturally arises in the low-energy limit
of string theories. In this case, the existence of
black-hole solutions will shed a new light on the semi-classical
effects of string theory on black holes. These prospects
presented here are left for possible future investigations. In the
same spirit, we can also consider a general $p$-form coupled to a
dilaton field as those that occur in standard supergravity theories.
The advantage of these models is that the presence of the dilaton
field permits to extend the notion of the electric-magnetic duality,
and hence the existence of magnetic solutions is tied to the
electric solutions.

%======================================%
%<<<<<<<<<< ACKNOWLEDGEMENT >>>>>>>>>>>%
%======================================%
\acknowledgments 
The authors thank Jorge Zanelli and Ricardo Troncoso for comments.
HM would like to thank Masato Nozawa for many helpful comments. 
HM would like also to thank the Max Planck Institute for Gravitational Physics (Albert Einstein Institute) and the Yukawa Institute of Theoretical Physics for hospitality and support.
This work has been partially funded by the
following Fondecyt grants: 1100328 (HM); 1090368 (MH); and  1085322, 1095098, 1100755 (CM). 
This work was also partly supported by the JSPS Grant-in-Aid for Scientific Research (A) (22244030).
The Centro de Estudios
Cient\'{\i}ficos (CECS) is funded by the Chilean Government through
the Millennium Science Initiative and the Centers of Excellence Base
Financing Program of Conicyt, and Conicyt grant ``Southern
Theoretical Physics Laboratory" ACT-91. CECS is also supported by a group of
private companies which at present includes Antofagasta Minerals,
Arauco, Empresas CMPC, Indura, Naviera Ultragas, and Telef\'{o}nica
del Sur. CIN is funded by Conicyt and the Gobierno Regional de Los
R\'{\i}os.

\end{document}